\documentclass[twocolumn]{aa} 
 
\usepackage{natbib}      
\bibpunct{(}{)}{;}{a}{}{,} 

\usepackage{txfonts}
\usepackage{caption}
\usepackage{subcaption}
\usepackage{graphicx}
\usepackage{bm}
\usepackage[utf8]{inputenc}
\usepackage{float}
\usepackage{stfloats}
\usepackage{color} 
\usepackage{xcolor}

\usepackage{hyperref}
\hypersetup{
colorlinks=true,
linkcolor=blue,
urlcolor=blue,
citecolor=blue,
}
\begin{document} 

\newcommand{\twolin}{\leavevmode \\ \leavevmode \\}
\newcommand{\trelin}{\leavevmode \\ \leavevmode \\ \leavevmode \\}
\newcommand{\graphics}{plots_finalfinal}
\newcommand{\graphicsmore}{plots_finalfinal}
\newcommand{\graphicsmoree}{plots_finalfinal}

\newcommand{\gf}{plots_finalfinal}
\newcommand{\gfstr}{plots_finalfinal/streams}
\newcommand{\cmg}{\rm\, cm^2 \, g^{-1}} 
\newcommand{\gcm}{\rm\, g \, cm^{-3}} 

   \title{On the structure and mass delivery towards circumplanetary discs}
	 \titlerunning{Structure and mass delivery towards CPDs}
	
   \author{Matth\"aus Schulik,
          \inst{1}
					Anders Johansen,
					\inst{1}
          Bertram Bitsch, 
					\inst{2}
					Elena Lega
					\inst{3}
					Michiel Lambrechts \inst{1}}
			
	\authorrunning{Schulik et al.}

   \institute{Lund Observatory, Box 43, S\"olvegatan 27, SE-22100 Lund, Sweden \\
              \email{schulik@astro.lu.se}
						\and
						Max-Planck Institut f\"ur Astronomie, K\"onigsstuhl 17, 69117 Heidelberg, Germany
						\and
             Laboratoire Lagrange, UMR7293, Universit\'e de la C\^ote d'Azur, Boulevard de la Observatoire, 06304 Nice Cedex 4, France
             }

   \date{Received ...}

\abstract{Circumplanetary discs (CPDs) form around young gas giants and are thought to be the sites of moon formation as well as an intermediate reservoir of gas that feeds the growth of the gas giant. 
How the physical properties of such CPDs are affected by the planetary mass and the overall opacity is relatively poorly understood. In order to clarify this, we use the global radiation hydrodynamics code FARGOCA, with a grid structure that allows resolving the planetary gravitational potential sufficiently well for a CPD to form.
We then study the gas flows and density/temperature structures that emerge as a function of planet mass, opacity and potential depth. Our results indicate interesting structure formation for Jupiter-mass planets at low opacities, which we subsequently analyse in detail.
Using an opacity level that is 100 times lower than that of ISM dust, our Jupiter-mass protoplanet features an envelope that is sufficiently cold for a CPD to form, and a free-fall region separating the CPD and the circumstellar disc emerges. Interestingly, this free-fall region appears to be a result of supersonic erosion of outer envelope material, as opposed to the static structure formation that one would expect at low opacities. Our analysis reveals that the planetary spiral arms seem to pose a significant pressure barrier that needs to be overcome through radiative cooling in order for gas to free-fall onto the CPD. 
The circulation inside the CPD is near-keplerian and modified by the presence of CPD spiral arms.
The same is true when we deepen the planetary potential depth, which in turn increases the planetary luminosity, quenches the formation of a free-fall region and decreases the rotation speed of the envelope by 10\%.
For high opacities we recover results from the literature, finding an essentially featureless hot envelope.
With this work, we demonstrate the first simulation and analysis of a complete detachment process of a protoplanet from its parent disc in a 3D radiation hydrodynamics setting.}


   \keywords{giant planet formation --
                simulations --
                radiation hydrodynamics
               }

   \maketitle
%

\section{Introduction}
\label{sec:intro}
%
Circumplanetary discs (CPDs) are rotationally supported discs, consisting of gas and dust, that are thought to form around massive host protoplanets. This formation process involves gas from the parent circumstellar disc (CSD), in which the planet is formed, being accreted into the Hill sphere of the planet. Inside the Hill sphere, the gas is unable to finalize its fall onto the planet as the gas cannot easily get rid of its angular momentum with respect to the planet, thus forming a disc \citep{tanigawa2012}. The recent discovery of a massive protoplanet with an associated mass of possibly circumplanetary material in PDS70 \citep{Keppler2018,haffert2019,isella2019} has stirred renewed interest into the properties of this class of objects. 

CPDs have been long recognized as potential sites for the formation of regular moon populations around giant planets \citep{canup2006, shibaike2019, ronnet2020}. Because they are directly coupled to the global CSD gas flows, CPDs are thought to be possible bottlenecks for further growth of giant planets. 
Those global flows can exhibit supersonic spiral arm shocks as the planet orbits relative to the CPD medium \citep{goldreich1978}, and once the planet grows massive enough those shocks perturb the global flows sufficiently to open gaps in the gas distribution \citep{goldreich1980}. Those gaps are thought to further limit the accretion rates into the planetary Hill spheres and in turn also to regulate CPD and planet growth \citep{dangelo2008}. 

On the other hand, radiative feedback from the growing planet and CPD into the CSD is inevitable, with potentially observable consequences for the CSD chemistry \citep{cleeves2015}. 

The study of such a triplet system of planet -- CPD -- CSD must therefore be performed numerically, in order to access all available physical information.

Because of the inherent complexity of simulating this problem, hydrodynamical simulations are usually either limited in the number of CSD orbits they can compute, or in the physics that is being computed. 
In this endeavour, a number of interesting milestones exist:

\cite{dangelo2003} used the system of 2D Euler equations with a cooling prescription and found various CPD structures with spirals in them.
Contrary to this work, \cite{ayliffe2009b} found no CPD spiral arms in a global, 3D radiation hydrodynamics setting, arguing that they disappear due to the increase in degrees of freedom for the flow in 3D. 
\cite{tanigawa2012} found a disk at keplerian rotation in their isothermal 3D simulation. \cite{gressel2013} showed in a comparison of global isothermal, adiabatic-hydrodynamic and MHD simulations that isothermal simulations favour highly-keplerian disc rotation and strong flattening of the CPDs. \cite{zhu2016} found again spiral arms in highly resolved radial-azimuthal 2D runs with a cooling prescription as function of optical depth, identified them to be triggered by the tidal action from the host star and measured accretion rates onto the planet through the CPD. The same study also investigated the effects of the optical depth on the CPD properties in detail by changing the CPD mass. An interesting result from their approach of varying the CPD mass is that the temperature changes with the optical depth. The temperature in turn controls the spiral arms attack angle, which is in their study responsible for the ongoing accretion of the CPD into the planet. Subsequently, \cite{judith2016} clarified in 3D radiation hydrodynamics runs that the occurrence of CPDs, as opposed to spherical envelopes around giant planets, is linked to their temperature, i.e. under otherwise identical conditions a hot envelope will collapse into a disk when cooled artificially. In total, those studies made clear that the exact accretion rates, rotational and structural properties of simulated CPDs depend on the numerical framework and the treatment of thermodynamics.

\begin{table}
\label{tab:simdata}
\centering
\caption{Simulation runs used in this paper. The relation of the runs with the science questions posed in this work are explained in Sec. \ref{sec:intro} and parameters are explained in Sec. \ref{sec:methods}.}
\begin{tabular}{c c c c c}        
\hline\hline                 
Label & $m_{\rm P}/m_{\rm Jup}$ & $m_{\rm P}/m_{\rm \oplus}$ & $\tilde{r}_{\rm s}$ & Opacities*  \\    
\hline
CPD occurrence \\
m1 & 0.06 & 20 &0.1 &  1.0/0.01  \\
m2 & 0.2 & 60  &0.1 &  0.01  \\
m3 & 0.3 & 90 &0.1 &  1.0/0.01 \\
m4  & 0.4 & 120 &0.1&  1.0/0.01  \\
m5  & 0.75& 240 &0.1&  1.0/0.01  \\
\hline
Main simulations \\
C1 ("nominal")  & 1 & 360 & 0.1 & 0.01 \\      
H1 ("high opacity")  & 1 & 360 & 0.1 & 1.0 \\
H2 ("deep") & 1 & 360 & 0.025 & 0.01 \\
C2 ("Bell \& Lin") & 1 & 360 & 0.1 & 0.01\\
\hline                                   
\label{tab:simdata2}
	\end{tabular}
	\\[5pt]
	\caption*{(*) Opacities used: numbers denote constant opacities $\kappa$ for all simulations in $\rm cm^2\,g^{-1}$, except for simulation C2, which uses Bell \& Lin opacities reduced by the factor in the table. All simulation runs have an unperturbed CSD surface density $\Sigma_0=100\; \rm g\, cm^{-2}$ at the planets position. Simulation letters refer to the purpose or expected simulation outcome: m="mass survey", C="cold", H="hot"}
\end{table}
%
%
%

Additionally, in \cite{schulik2019} (hereafter S19), we showed that underresolving the central planetary region raises temperatures (and entropies) to incorrect levels. Furthermore, reducing the opacities of the planetary envelope was demonstrated to aid the flattening of envelopes surrounding Saturn-mass planets significantly.
Thus, there is a clear need to re-visit past results about CPDs in the framework of 3D global, radiation hydrodynamics simulations with sufficient numerical resolution and a variation of opacities for Jupiter-mass planets. 
%

With this motivation, we can ask and address a number of science questions, that are related to our two sets of simulations listed in Table \ref{tab:simdata2} as follows:
\begin{itemize}
    \item At which protoplanet masses do CPDs appear? We address this by quantifying the flatness and state of rotation of circumplanetary envelopes for a number of protoplanets with increasing masses in runs m1-m5, and C1 / H1.
    \item What are the masses and accretion rates of CPDs relative to those of the planet?
    Due to the coupling of gas flows from the CSD into the planetary Hill sphere, CPDs will to some degree co-evolve with their host planet. This process will be particularly important if the CPD is accreting onto the planet and at the same time is replenished from the CSD. Earlier authors have often assumed that the mass accreted by the planet is a fraction of the mass entering either the Bondi or the Hill radius \citep{dangelo2008, owen2016}, but a more recent study \citep{lambrechts2019} shows that the mass flux through the Hill sphere and the actual accretion rate can differ by a factor of up to 100, with this number approaching $\sim$0.1 when the planetary mass reaches $\sim$$1 M_J$. We use runs C1/C2 and H1/H2 to extend this idea and measure the mass and accretion rates of the CPDs and compare them to those of the planets.
    
    \item What is the influence of the variation of dust opacity on properties of the CPD? Is there an influence from the CSD on the accretion process? Discs at different ages are able to provide differing amounts of dust to already formed protoplanetary envelopes \citep{brauer2008, birnstiel2012}. However, the full time evolution of CPDs in 3D radiative hydrodynamics simulations is currently not accessible on modern hardware. This problem is even more severe if one would consider the evolution of the full disc, including stellar irradiation and the evolution of the size distribution of the dust. Thus we take a simplified approach and compare a dust-rich (run H1) and dust-poor (run C1) scenario. We also perform two support parameter simulations, investigating deep planetary potentials (run H2) and more realistic, non-constant opacities, given by \cite{belllin1994} (run C2). 
\end{itemize}
We structure this paper as follows in order to address those questions. In Section \ref{sec:methods} we describe our methods, the used set of simulations and justify parameter choices. Section \ref{sec:results_occurence} describes how the envelopes of protoplanets of varying masses and envelope opacities flatten and rotate. Section \ref{sec:results_nominal} describes and analyses our nominal simulation run, which presents most of the novel and interesting physics features. Subsequently, Section \ref{sec:results_comparison} discusses how those features change when using different simulation parameters and establishes the robustness of our nominal results. That section also presents results for CPD masses, mass profiles and accretion rates, together with reflections upon those.

\section{Methods}
\label{sec:methods}

The methods used to address the science goals are mostly identical to our previous work, \cite{schulik2019} (S19). We briefly reiterate them here and point out major changes. 

\subsection{Physical model and parameters}

We use the code FARGOCA \citep{bitsch2014,lega2014} to solve the equations of viscous hydrodynamics coupled with an evolution equation for the zeroth moment of the photon intensity, which is the mean photon energy density equivalent to a separate photon temperature. The hydrodynamic advection problem is solved using the FARGO algorithm \citep{masset2000}, while momentum transport evolves also via the full viscous tensor through the employment of a classical Strang-splitting scheme. The radiation transport is solved with the help of flux-limited diffusion \citep{levermore1981} and we set $\kappa_{\rm R}=\kappa_{\rm P}\equiv \kappa$ for reasons of simplicity.  
The equation of state for the gas is adiabatic with a constant adiabatic index of $\gamma=1.4$. The mean molecular weight is that of a solar hydrogen-helium mixture, which is $\mu =2.35 \rm \,g/mole$. Viscous momentum transport is computed through the full hydrodynamic stress tensor multiplied with a constant physical viscosity of $\nu = 10^{-15}\rm \,cm^2/s$ that corresponds roughly to $\alpha \approx 10^{-2}$ at our disc temperatures. \cite{fujii2017} found that CPDs will not be hot enough to be ionized, therefore we do not include the effects of magnetohydrodynamics into our simulations.

The main parameters of interest for our work are the planetary masses, the opacities and the gravitational smoothing lengths. Their variations and corresponding simulation labels are listed in Table \ref{tab:simdata2}.

The planetary masses used are $20\, m_{\oplus}$ up to $360\, m_{\oplus}$; the latter is just slightly above 
$\approx$$1\, m_{\rm Jup}$. We use two different constant opacities, as listed in Table \ref{tab:simdata2}, to scan the parameter space of CPD properties. High opacity runs 
($\kappa=1.0 \rm\, cm^2/g$) serve to connect to previous work by \cite{lambrechts2017} (for the 20$\, m_{\oplus}$ case) as well as \cite{judith2017} and \cite{lambrechts2019} (for the Jupiter-mass planets). 

Quantities denoted with a tilde, such as $\tilde{x} = x/x_{\rm H}$ have numerical values indicating the Hill radius fraction. This is similar to the notation used in \cite{tanigawa2012}, where all quantities were normalized with the pressure scale height $h$, but as they used $r_{\rm H} = h$, the locations of circulation features in our plots can be directly compared to those in that work.

Gravity in our simulation domain is determined by contributions from both the star and the planet. The planetary gravity consists only of a gravitational potential with an inner cut-off defined by the smoothing length $r_{\rm s}$ \citep{klahrkley2006}. This is a key numerical parameter in our work. All our simulations are performed with well-resolved smoothing lengths (10 cells per $r_{\rm s}$, as defined in S19) and a nominal potential depth of $\tilde{r}_{\rm s}=0.1$. This value has been chosen after reviewing the literature on CPDs, indicating CPD outer edges at around $\tilde{r}=0.3-0.5$.
In one case, that of our most expensive simulation (run H2), we go as deep as $\tilde{r}_{\rm s}=0.025$, which corresponds to a physical size of $r = 12 \,r_{\rm Jup}$ at our orbital distance of $5.2\rm AU$.

The independent evolution of gas internal energy and mean photon energy density allows the code to find regions where those two quantities are either at equilibrium or are uncoupled from each other. The former is the case in dense, optically thick media while the latter is the case for very tenuous regions of the simulation domain. The coupling between gas and photon energy is calculated using the Planck-mean opacity, while the diffusion/escape of photons into space is calculated using the Rosseland-mean opacity; both are taken as equal in this work and we subsequently only refer to them as the opacity. Then, we use either different values of constant opacities $\kappa$, or those from \cite{belllin1994}, scaled by a factor, which we list as $\epsilon$ in the last column  in Table \ref{tab:simdata2}. Those opacity values are compatible with the upper and lower limits of opacities suggested in \cite{mordasini2014} which correspond to a typical factor of $10^{2}$ to $10^{4}$ reduction compared to ISM dust opacities. The exact equations we solve for the radiative transfer can be found in S19. Planetary gaps are generated via the same process as in the previous paper and are evolved for 400 orbits before the resolution is increased to resolve the CPD.

\subsection{Numerical setup and grid refinement}

We work in a global, spherical coordinate system that is centered onto the star and label the independent coordinate as $[r,\theta,\phi]$. We also use the planetocentric coordinates $[x,y,z]=[(r-1)\, \cos{\theta} \cos{\phi},r\, \sin{\theta} \cos{\phi}, r \, \sin{\phi} ]$, so that the planet will be at $[x_p,y_p,z_p]=[0,0,0]$ and we will often refer only to the planetocentric coordinates for simplicity.

Boundary condition data are generated as in S19 and before in \cite{kley2009, lega2014, bitsch2014} via the usage of three distinct numerical steps. The first, generates a radiative, radial-vertical disc equilibrium in 2D with a sufficiently wide radial extent. In the second step, the gap formation step, the data from the first step is run in low, equidistant resolution in 3D for 400 orbits, which has been previously shown in S19 to be deep enough for significant gap depths. Finally, the third step takes a limited radial extent, as explained below, of the global disc and runs it in high resolution around the planet. 

The simulation domain in all our simulation runs is $\theta \in [-\pi,\pi]$, $\phi \in [81, 90]^{\circ}$ for azimuth and colatitude. Boundary conditions in azimuth are periodic for all variables. In the colatitude, we use a half-disc, i.e. reflective boundaries at $\phi=90^{\circ}$ for all variables. Reflective boundaries are used for hydrodynamic variables at $\phi=81^{\circ}$, and open boundaries for the radiative energy at $\phi=81^{\circ}$. The radial extent of our simulation domains is adjusted depending on the planet mass. For $20\,m_{\rm \oplus}<m_{\rm p}<120\, m_{\rm \oplus}$ it is sufficient to run with $r \in [0.7,1.3] \times 5.2\,\rm AU$. However, once planetary gaps become deeper for higher masses, the gap width also increases. Hence we run the massive protoplanets with $m_p > 120 \, m_{\rm \oplus}$ with $r \in [0.4,1.6] \times 5.2\,\rm AU$. Using a too narrow radial simulation for the high mass planets results in artificial gap edge instabilities that feed large surges of gas into the planet. Those instabilities disappear once running with the larger radial extent. 

The spacing of the simulation grid in radial, azimuthal and colatitudinal extent is of the form $d\Xi \propto 1- g(\Xi-\Xi_0)$, where $g(\Xi)$ is a gaussian function, $\Xi$ represents any of the three spatial directions ${r,\theta,\phi}$ in our spherical coordinate system and $\Xi_{\rm 0}$ is the planetary position for any of the three coordinates. 
This type of grid results in a highly-resolved, quasi-cartesian grid in the direct vicinity of the planet. We design the three grids in such a way that the increase in resolution element size has the same value for all three spatial dimensions, in order to not introduce artificial effects of overresolution in any one direction. In order to minimize computation time, the resolution gradients are stretched to a reasonable maximum that leaves the horseshoe-orbits unchanged and allows radial density gradients in the gap to converge numerically. 

The choice of the minimum resolution element $d\Xi_{\rm min}$, i.e. the element with the best resolution directly adjacent to the planet and maximum of $g(\Xi)$, is given by the requirement for the gravitational smoothing length to be resolved by 10 cells and is hence determined after the choice of $r_{\rm s}$.
This requirement originates from our previous studies focused on accretion rates (S19), but we find that well-resolving $r_{\rm s}$ is equally important for the structure of CPDs in order to avoid artificial overheating and artificial flattening of the planetary envelopes. For the nominal simulation run this results in a minimum resolution element of size $dx_{\rm min}=7 \times 10^{-4}$ (or $7.5 \, r_{\rm Jup}$ )
and our simulation run H2 with the deep potential, has $dx_{\rm min}=1.7 \times 10^{-4}$ ( or $1.9\, r_{\rm Jup}$).

\section{Results - Occurence of CPDs with increasing planetary mass and the influence of opacity}
\label{sec:results_occurence}

We now turn to present the results from our mass survey. In general we speak about planetary envelopes when referring to the gas surrounding the smoothing length, as this nomenclature remains agnostic towards the existence of a CPD.

\subsection{Measuring the flatness of envelopes}

In order to assess the structure of the planetary envelopes, one could pursue the idea of measuring the envelope aspect ratio $H/r$. However, $H/r$ does not inform about the keplerian rotation support of the envelope. This is because without an assumption about the ratio of pressure to centrifugal support, the aspect ratio can be only recast into the form 

\begin{align}
\frac{H}{r} = {\left( \frac{c_s}{v_k} \right)}^2,    
\end{align}
which is not $v_{\theta}/v_{\rm k}$.
Hence, in order to measure the flatness of a disc we define the flatness parameter at distance $d$ through the cylindrically averaged 2D-density $\bar{\rho}(r,z)$
\begin{align}
\xi(d) \equiv \frac{\bar{\rho}(0,d)}{\bar{\rho}(d,0)}.
\end{align}
This definition is a simple measure for the asymmetry between the density profiles in vertical versus the midplane directions and thus an indicator for 'disciness'. We use this parameter $\xi$ additionally to the quantity $v_{\theta}/v_k$ to assess the properties of planetary envelopes with increasing mass, for the simulation sets $m1-m5$, C1 and H1.

\begin{figure}
	\centering
	\includegraphics[width=0.5\textwidth]{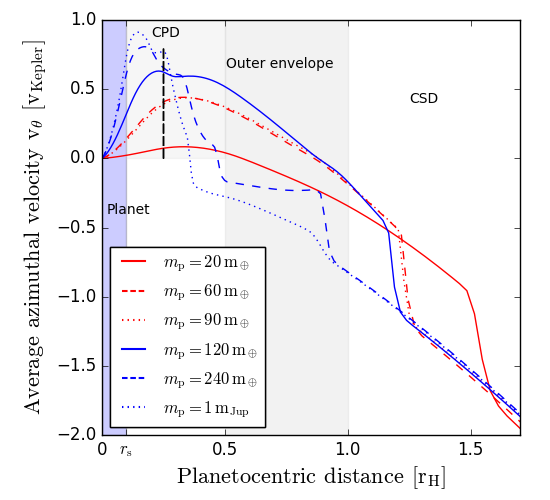}

\caption{Rotation profiles in the envelopes of planets of varying masses. We define the regions in the surroundings of the planet according to the rotational state of the envelope. Note that the region labelled as CPD, is only appropriately named for masses of $m_p \geq 240 \rm m_{\rm \oplus}$. The strict transition into the keplerian shear of the CSD happens at $v_{\theta}/v_{\rm Kepler} = -1$, as can be seen from the slope turnover in the rotation curves. The vertical, black dashed curve denotes $r=0.3r_{\rm H}$, where rotation and density asymmetry values are measured and plotted in Fig. \ref{fig:discs_appearing}. }
	\label{fig:discs_rotation}
\end{figure}

\subsection{Envelope flatness}

We first present the simulation data which we used to inform our later decisions on identifying key interesting simulation parameters to be studied in detail.

We investigated the rotational state of the midplane around planets of various masses at a time of 5 orbits after the start of our simulation runs. This allows the planetary envelope enough time to settle into rotational equilibrium, to the degree that the counteracting pressure gradients permit this. This is also enough time to establish an accretion equilibrium between gas flowing from the envelope onto the planet and accreted high-angular momentum gas replenishing gas in the envelope.

\begin{figure}
	\centering
	\includegraphics[width=0.5\textwidth]{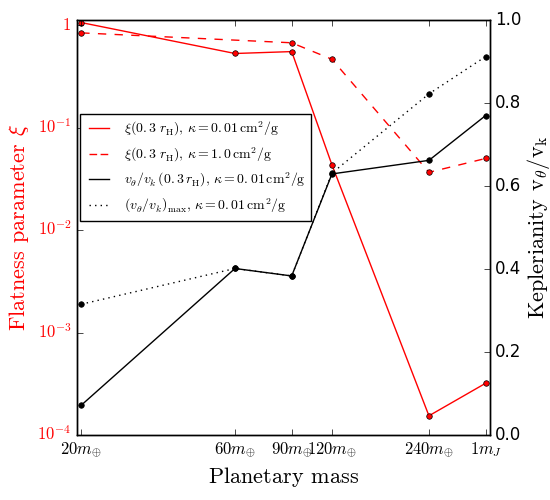}

\caption{Occurrence of CPDs as measured by flatness parameter and keplerianity of planetary envelopes of different masses. Simulation data is taken in steady-state after 5 orbits runtime. The maximum value of $v_{\theta}/v_{\rm K}$ is also indicated in order to indicate fluctuations along one orbit. While the keplerian rotation fraction of the protoplanetary envelopes is rising quasi-linearly with the mass, the flatness shows a sharp decrease at 120$\,\rm m_{\oplus}$, for both values of the constant opacities. This shows that the infall and accretion of angular momentum in the midplane follow a straightfoward relation with the potential that they are accreted into. On the other hand, the vertical cooling responsible for the envelope flattening plays an independent role and prevents CPD formation at high opacity, even if the necessary potential depth is given.}
	\label{fig:discs_appearing}
\end{figure}

We define the distinction between planet, circumplanetary disc and envelope as follows. The planet is all material inside the sphere of radius defined by the smoothing length, i.e. $r=r_{\rm s}$, centered on the centroid coordinates of the planetary potential, i.e. $(r_{\rm p},\theta_{\rm p},\phi_{\rm p})=(1,0,\pi/2)$. We emphasize that this is neither a real planetary surface, nor a planetary interior.
The CPD we define as the region inside of $0.5 r_{\rm H}$ with prograde/positive rotation.
This definition for the CPD excludes the region from $0.5  < \tilde{r} <1.0$, as this is a region into which horseshoe orbits penetrate (seen in Fig. \ref{fig:discs_rotation} as regions of retrograde/negative rotation.), which we call the \textit{outer envelope}. The sum of CPD and outer envelope we simply call the \textit{envelope}. Hence, material entering the Hill sphere will enter the envelope through the outer envelope.

\begin{figure*}
\hspace*{-0.5cm}	
  \begin{subfigure}{0.43\textwidth} 
   \centering
   \includegraphics[width=1.1\textwidth]{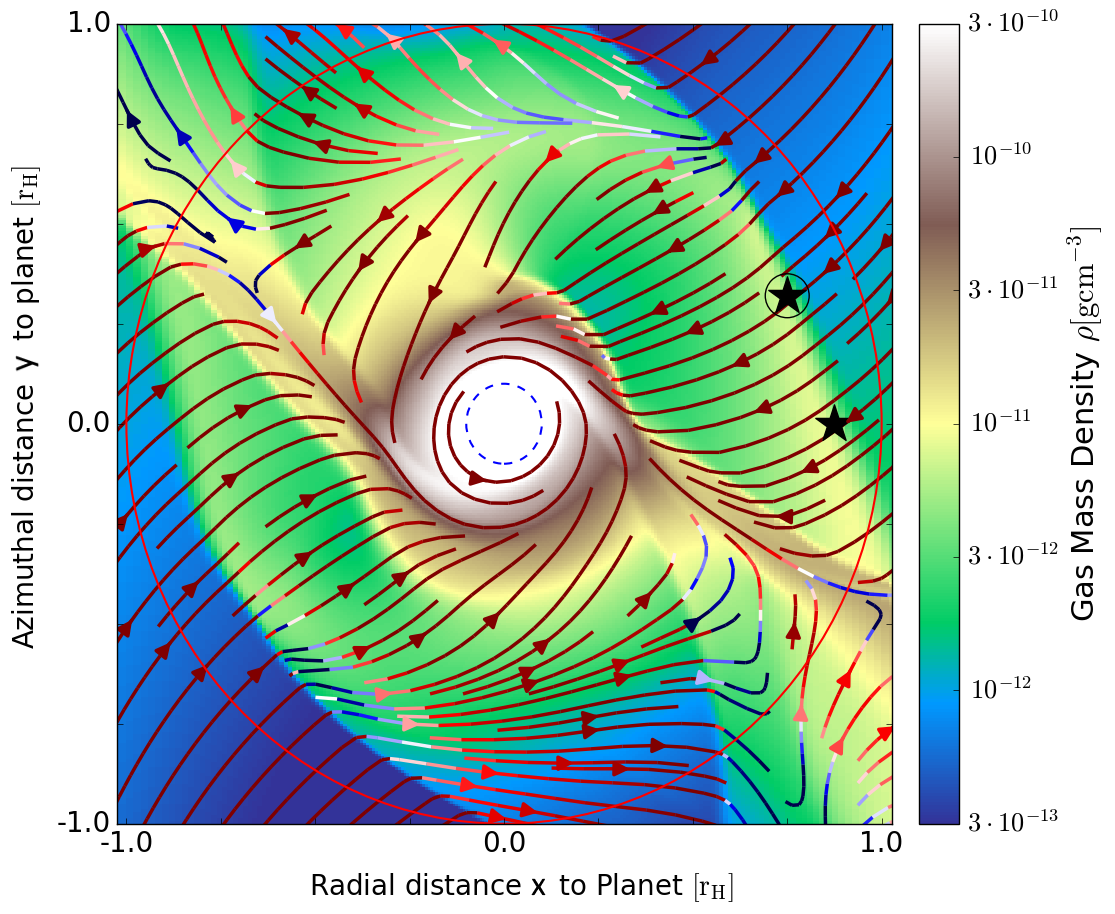}
	\end{subfigure}%
	\hspace*{+1.50cm}
	\begin{subfigure}{0.48\textwidth} 
   \centering
   \includegraphics[width=1.1\textwidth]{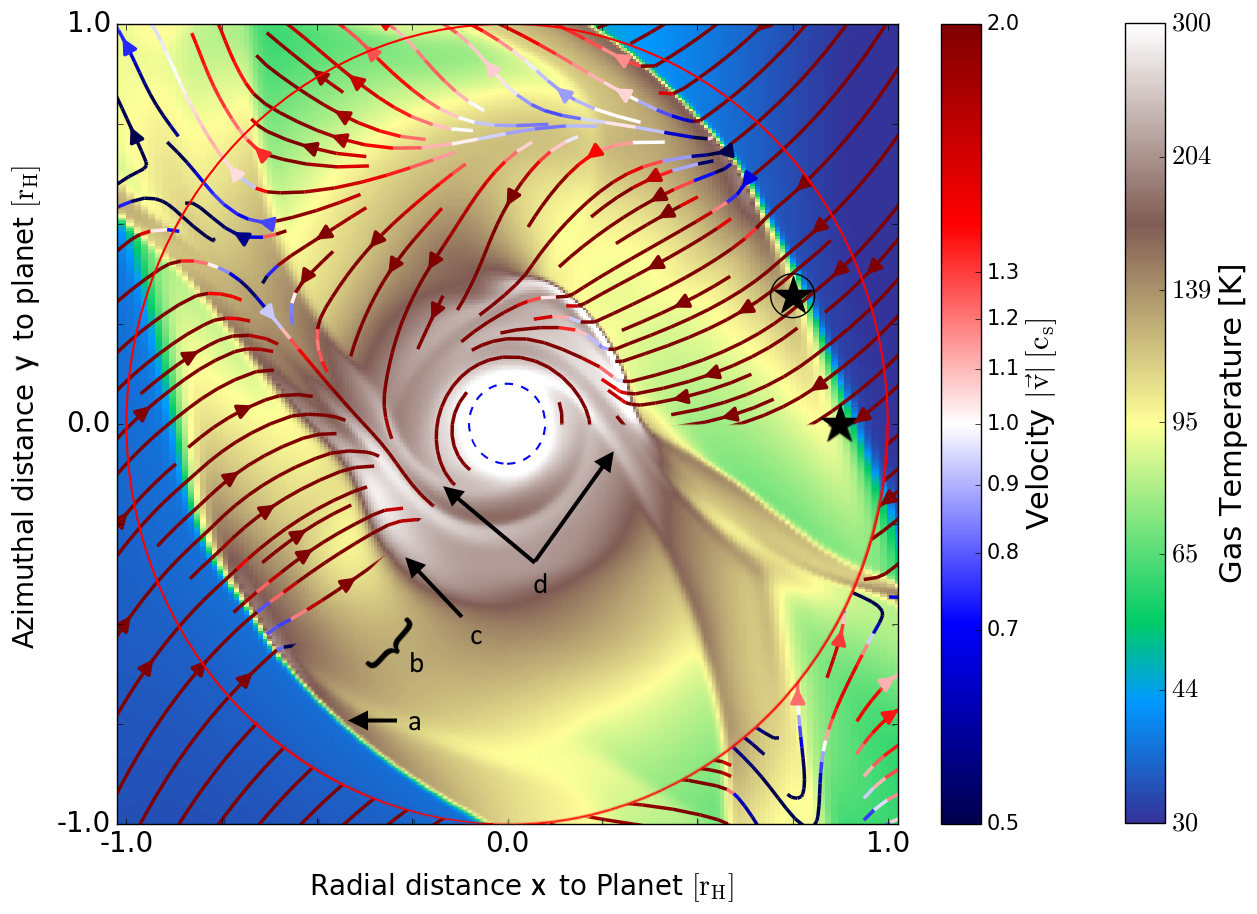}
\end{subfigure} 
\caption{Overview of flow structures in the mid-plane of the nominal simulation run C1 after reaching steady state at orbit 5. Density (\textit{Left}) and temperature (\textit{Right}) values in the midplane are shown along with the gas streamlines coloured according to Mach number. While streamlines in all generality do not coincide with the gas motion, they do so in steady-state. Gas coming from the circumstellar disc encounters the spiral arms in the midplane, which are heavily modified by the ongoing accretion process and non-isothermal modification of the planetary Hill sphere. The rotationally supported disc is formed mainly between $\tilde{r} = 0.1$ and $\tilde{r} = 0.4$. This disc contains its own structures, such as a midplane accretion shock from the supersonically infalling gas and smaller CPD spiral arm shocks. Arrows target the features, not specific points. The letter-labels refer to CSD spiral arm (a), free-fall region (b), CPD accretion shock (c)  and CPD spiral arm (d). The black stars denote the positions where streamlines of interest cut through the spiral arm shock. They have constant positions in 3D, and will serve as orientation points when investigating the vertical direction. Note the highly supersonic flows which remain supersonic after encountering the spiral arm shock. This is due to insufficient static shock pressure, which is advecting  the spiral arm radially inwards. The spiral arm mass is replenished from the vertical direction.}
\label{fig:c1_midplane_dens_temp}

\vspace*{+0.20cm}
\end{figure*}

\begin{figure}   
	\centering
	\includegraphics[width=0.45\textwidth]{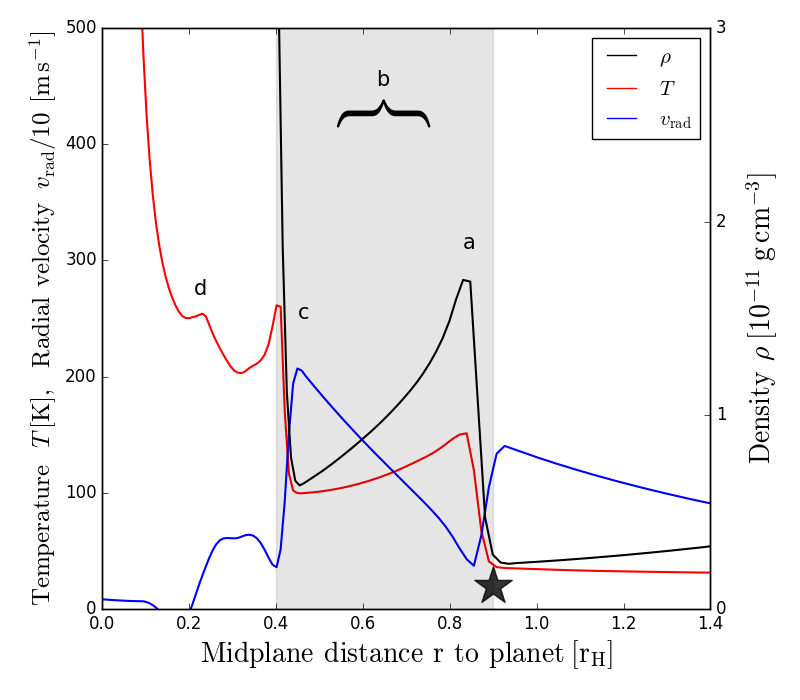}
         \label{fig:c1_midplane_profiles}

\caption{Overview of slices through the structures for the non-circled star in the nominal run C1 at orbit 5 for the midplane. The letter labels correspond to the same featureas as in Fig. \ref{fig:c1_midplane_dens_temp}, only on the opposite side. In the midplane, the most prominent feature is the free-fall region (marked with b and braces, \textit{grey area}), just after the spiral arm shock (a). The free-fall is notable as the density profile decreases as the radial velocity accelerates. The free-fall is terminated when the gas hits the CPD at the left edge of the grey area, seen as sharp increase in density and as CPD accretion shock (c). Note that we use a linear density scale here for emphasis. The CPD spiral arm is notable as temperature bump (d). The black star denotes the same position as in Fig. \ref{fig:c1_midplane_dens_temp}.}	\label{fig:c1_midplane_profiles}

\end{figure}   
\begin{figure}
   \centering
   \includegraphics[width=0.45\textwidth]{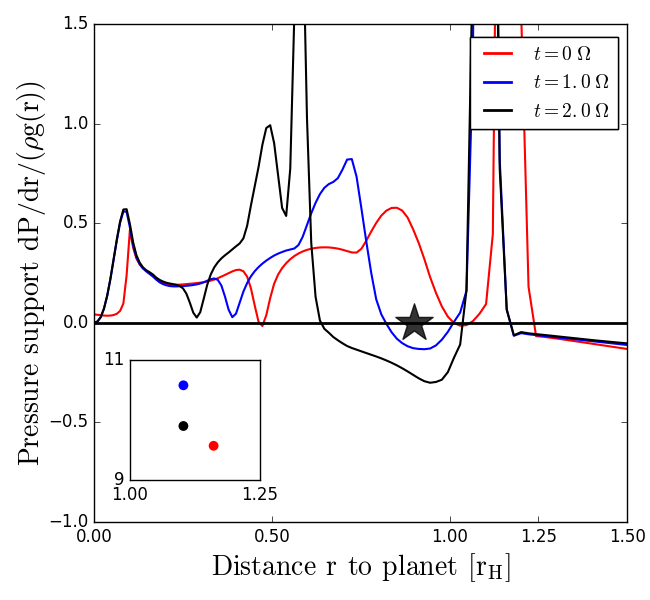}
\caption{Cuts through the pressure support vs. radius inside the Hill sphere for the same times as Fig. \ref{fig:ffr_evolution}. Of particular interest is the state of the simulation at $t=0\, \Omega$ just after the gap formation run. We find the envelope is in a state of latent imbalance, due to the low resolution in the gap formation run. From there, the CSD flows push into the envelope, until a new equilibrium is found. The inset shows the evolution of the static shock pressure support. The difference in position between the star and the shock positions at $t=0-2\Omega^{-1}$ showcase the slow evolution of the spiral arms between those times. The black star denotes the same position as in Fig. \ref{fig:c1_midplane_dens_temp}.}
         \label{fig:pressuresupport_time}
\end{figure}

For all planet masses we measure the rotation profiles in their envelopes and plot them in Fig. \ref{fig:discs_rotation} normalized to each planet's individual Hill radius. Between planetary masses of $20 \,\rm m_{\oplus}$ and $60\,\rm m_{\oplus}$ we see an important evolution of the rotational profile. A general steepening and retreat of the transition into the Keplerian shear of the disc is evident. From $60 \,\rm m_{\oplus}$ to $120\,\rm m_{\rm \oplus}$ there is only a relatively weak evolution in the envelopes. After a mass-doubling from $120\,\rm m_{\rm \oplus}$ towards $240\,\rm m_{\rm \oplus}$, a transition in the shape of the rotational profile occurs, developing a 'pedestial' consisting of relatively high Keplerian rotation in the CPD region for the $240\,\rm m_{\rm \oplus}$ and $360\,\rm m_{\rm \oplus}$ planets.

Since the parameter space of envelopes of planets around the classical, critical runaway core-mass have already been studied in detail in 3D radiation hydrodynamical settings \citep{ormel2015, lambrechts2017, kurokawa2018}, here we focus on the high-mass end of the planet evolution. A key parameter regulating the ability of a gaseous envelope to rotate is the opacity of the gas/dust - mixture. In order to assess its importance in forming rotating CPDs, our scan in planet masses was performed with two different constant opacities.

The flatness at $r=0.3\,r_{\rm H}$, being representative of the strongly rotating part of the envelope, and the fraction of keplerian rotation after 5 orbits in steady state are plotted in Fig. \ref{fig:discs_appearing}. It is evident, that particularly at higher masses, opacity plays an important role in setting the flatness and the rotational state of envelopes. The curve for the maximal values of $v_{\rm \theta}/v_{\rm Kepler}$ indicates the deviation of individual fluid elements along the orbit of radius $0.3\, r_{\rm H}$ and shows that the CPD for the Jupiter-mass planet with constant opacity of $\kappa=0.01\, \rm  cm^2/g$ rotates non-uniformly with individual fluid elements reaching up to 90\% of the keplerian value. 

Fig. \ref{fig:discs_rotation} shows clearly that protoplanets must reach approximately Jupiter-mass before a significant rotationally-supported CPD forms. We therefore focus in the next section on analysing the gas flow and density-temperature structure around the Jupiter-mass planet in the nominal simulation C1, which exhibits a richness of physical features. In Sec. \ref{sec:results_comparison} we continue to compare the C1 structure to the other Jupiter-mass simulations where we vary the physical and numerical parameters (runs C2-H2).

\begin{figure*} 

 \hspace*{+0.0cm}
	\centering
	\includegraphics[width=1.0\textwidth]{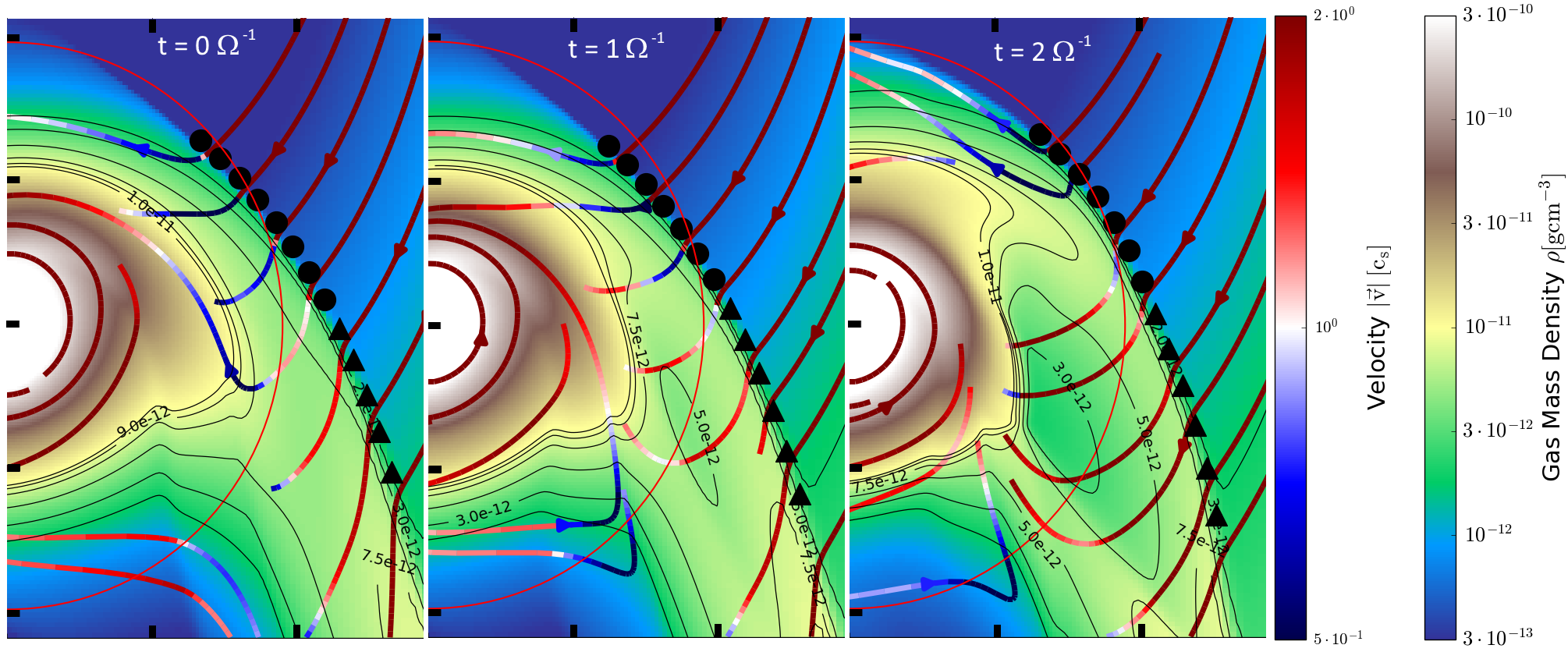}
      
\vspace*{+0.20cm}	
	\caption{Evolution of the FFR as seen in velocity as streamline colour, and density as background colour, with density contours to guide the eye. Snapshots are taken at an interval of $1.0$ $\Omega^{-1}$. Symbols along the spiral arm shock surface denote the pressure ratio $\zeta$, as defined in Eq. \ref{eq:pressure_ratios}. Filled circles denote approximate ram-pressure equilibrium, i.e. $\zeta \approx 1$, and filled triangles indicate $\zeta > 1$, i.e. the region where the streamlines can push past the shock. The $\zeta > 1$ streamlines remain supersonic after encountering the shock. Those post-shock supersonic streamlines evacuate the envelope efficiently, which helps accelerate the flow further (note the intensification of the red colour on the streamlines between the \textit{middle} and \textit{right} panels, which also corresponds to an increase of velocity in absolute numbers). Note also how the original spiral arm splits in two through this process. }
         \label{fig:ffr_evolution}
\end{figure*}

\section{Results - Gas flows and envelope structure in the nominal simulation run}
\label{sec:results_nominal}

This section presents the features that we find in the nominal simulation run C1. We first show and explain the simulation outcome in the midplane. We then do the same for the vertical direction, which exhibits more complex physics that couples into the midplane dynamics. For both midplane and the vertical direction, we describe three separate topics: the density and temperature structure of the envelope, mass delivery towards the CPD and circulation inside the CPD. 

\subsection{The midplane structure}
\label{sec:results_midplane_structure}

In Fig. \ref{fig:c1_midplane_dens_temp} we show the density and temperature structure together with the associated midplane flows. The innermost region around the planet forms an approximately spherically symmetric density and temperature distribution that extends beyond the smoothing length at $\tilde{r} = 0.1$ towards $\tilde{r} = 0.2$. A radial inflow exists which slows down as the pressure support increases. Between $\tilde{r} = 0.1$ and $\tilde{r} = 0.2$ the envelope's pressure support gradually fades until at $\tilde{r} = 0.2$ the centrifugal forces become significant enough to flatten out the envelope and let the gas rotate with $80-90\,\% \,v_{\rm K}$, see Fig. \ref{fig:discs_rotation}.

Between $\tilde{r} = 0.2$ and $\tilde{r} = 0.4$ we find the CPD proper, with time-independent spiral arm features that are evident in density as discontinuity and in temperature as local temperature maximum (marked as feature (d)). We term those spiral arms "CPD spirals" for the purposes of referencing them later in this work. The CPD spirals share a superficially similar morphology to those presented in \cite{zhu2016}. However due to our limitations in resolution and gravitational smoothing, we only see the outermost region of what Z16 are able to probe. Compared to that work, our spiral arms are fairly thick, which we attribute to our high viscosity, while their simulations are inviscid. We will discuss the three-dimensional effects of the spiral arms in the section about three-dimensional flows inside the CPD, Sec. \ref{sec:results_cpd_spiral}.

At $\tilde{r} \approx 0.4$ we find a midplane accretion shock, which results from the collision of the free-falling gas and the CPD (feature (c)). Its effects can be seen as a sharp outer rim in the midplane temperature at the CPD edge. Between $\tilde{r} = 0.4$ and $\tilde{r} = 0.9$ there is a region of free-falling CSD gas, which exhibits an inversion in pressure gradient (feature (b)). We subsequently name this region simply the free-fall region (FFR). Correspondingly, this region is evacuated relative to the CSD region and the density and temperature decrease inwards in the midplane as the radial velocity increases, which we plot in Fig. \ref{fig:c1_midplane_profiles}).

At even larger distances than $\tilde{r} \approx 0.9$ the spiral arms shock (feature (a) in Fig. \ref{fig:c1_midplane_dens_temp}), decelerate and funnel material into the FFR, a similarity to Bondi-Hoyle type accretion, as opposed to Kelvin-Helmholtz accretion that is regulated by cooling and contraction of the envelope. This interaction between the CSD spiral arm shocks, the accretion flow and the physical origin of the FFR will be further discussed shortly in Sec. \ref{sec:results_vertical_structure}.

Moving to the largest radial distances of interest, between $0.9 <\tilde{r} \lessapprox 6$, the planetary gap in the CSD becomes optically thin. This implies that photons emitted from the planet do not heat its immediate vicinity, but are only re-absorbed at the gap edge, keeping the gas flowing into the Hill sphere cool at $\approx30$K.

\subsection{Free-fall onto the CPD in the midplane}
\label{sec:midplane_freefall}

We now turn to a more detailed investigation into the origin of the free-fall region (feature (b)). A profile of the state of the hydrodynamic variables after the opening of the free-fall region can be seen in Fig. \ref{fig:c1_midplane_profiles}. In order to investigate the opening process of the FFR, we use the data from three snapshots in time, taken at $t=0,1,2\,\Omega^{-1}$. For those times, we plot the pressure gradient normalized to the local gravity, in the same slice through the non-circled star from Fig. \ref{fig:c1_midplane_dens_temp}, which we present in Fig \ref{fig:pressuresupport_time}.

There, it is evident that the initial pressure support in the outer envelope ($\tilde{r}\approx 1$) is near zero. This allows the circumstellar flows to penetrate the envelope (seen in the inset, as the shock pressure evolves upwards). A region of negative pressure support, i.e. free-falling gas is subsequently established in the outer envelope.

We expand this analysis by plotting how the density evolves together with the streamline velocities in the midplane in Fig. \ref{fig:ffr_evolution}. There, it becomes evident that the opening process seems to commence at around 1$r_{\rm H}$ distance just behind the spiral arm, and later continues to empty out an entire region behind the spiral arm in a asymmetric fashion. We think that efficient advection is responsible for this process. 
Advection is naturally more efficient for faster gas flow, hence the erosion of gas that opens the free-fall region is facilitated when the gas remains supersonic after the shock. Remaining supersonic after the shock is only possible for the flow when the ram-pressure of the incoming CSD material overwhelms the shock ram-pressure. Hence, in order to analyse this process further, we define the ratio of total pressure pre-shock to static pressure post-shock as

\begin{align}
\zeta \equiv \frac{(p_{\rm dyn} + p_{\rm stat})_{\rm pre}}{ (p_{\rm stat})_{\rm post}} = \frac{( \rho u_{\perp}^2+ \rho \gamma c_s^2)_{\rm pre}}{(\rho \gamma c_s^2)_{\rm post}}, 
\label{eq:pressure_ratios}
\end{align}
where $u_{\perp}$ corresponds to the streamline-velocity component perpendicular to the shock surface. We stress at this point that nominator and denominator of $\zeta$ are generally independent of each other due to the supersonic nature of the inflow from the circumstellar disc. We use the definition of $\zeta$ to analyse the time-evolution that is plotted in Fig. \ref{fig:ffr_evolution}. There, streamlines that are decelerated to subsonic post-shock speeds, can be identified as having $\zeta<1$. Those that remain supersonic post-shock coincide with the area at the shock where $\zeta > 1$. We will continue using this quantity to greater extent in the vertical direction, but are able to already draw some conclusions from this.

\begin{figure}	
	\hspace*{+0.50cm}
   \centering
   \includegraphics[width=0.47\textwidth]{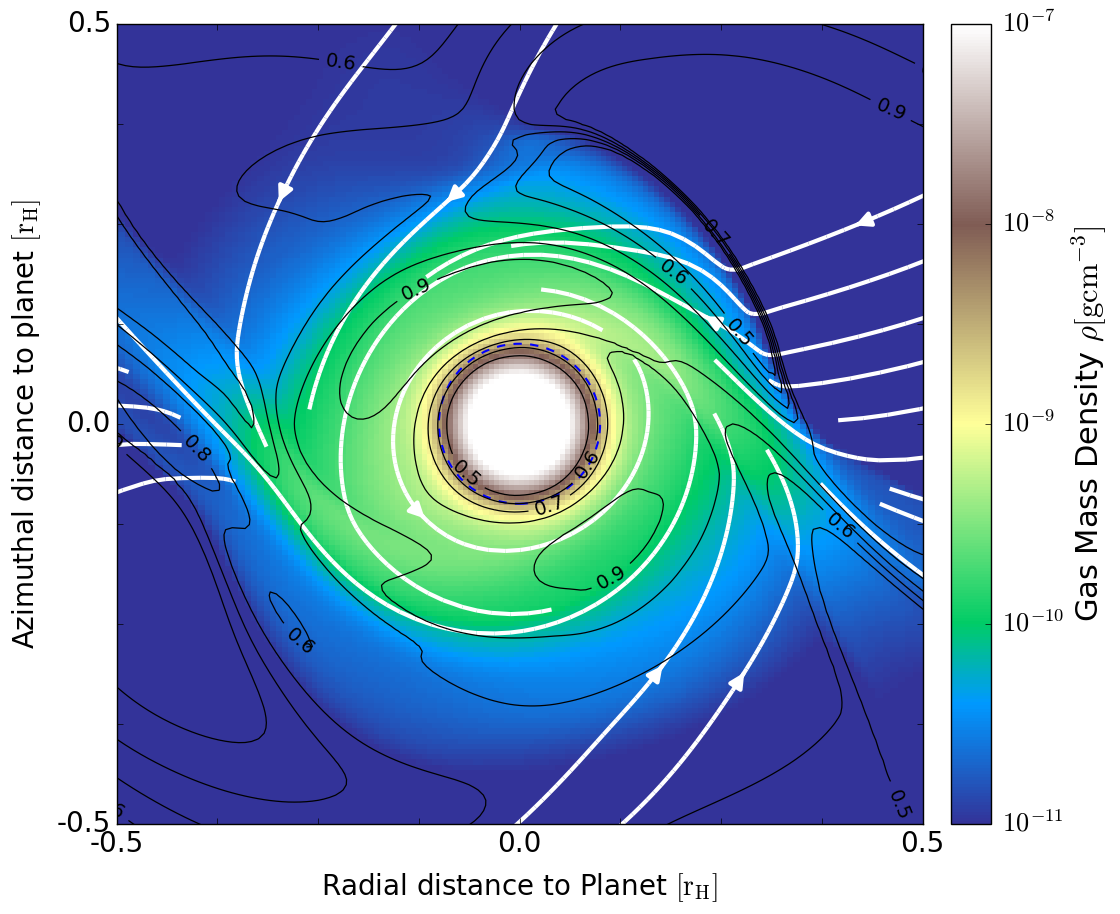}

\caption{Midplane rotational structure for the run C1 with $|\vec v|/v_{\rm Kep}$ as contours. After gas from the FFR enters the CPD it encounters one of the spiral arms and is shocked towards higher values of $|\vec v|/v_{\rm Kep}$. 
This results in a de-facto isolation between on one hand the infalling material into the FFR and on the other hand the planet in the midplane. This material nevertheless does find its way into the planet through a complex vertical circulation, as is clear from the streamlines in Fig. \ref{fig:c1_vertical_flows}. The blue, dashed circle marks the gravitational smoothing length.}
	\label{fig:rotiation_and_accretion}
\end{figure}

Once the incoming CSD flows possess sufficient dynamic pressure to overwhelm the static shock pressure, those flows notice the shock as if it were a speed bump, but not a dominant counteracting force, and continue at supersonic speeds. This helps enormously in transforming the initially massive spiral arm into a low-density region between two split, weaker arms. If the static shock pressure is high enough to act against the incoming flow, then the corresponding streamlines behave like a text-book shock, being decelerated according to the jump allowed by the Rankine-Hugoniot conditions in the planetary co-rotating rest-frame. One can also view this from a different perspective. At sufficiently high Mach-number, the post-shock material can always be advected with the flow. This would however impart a nonzero shock velocity. In the rest-frame of this shock, the post-shock material will always be subsonic, which we confirmed using a toy shock model. The remaining puzzling issue then is, how the spiral arm can remain in existence, as seen in Fig. \ref{fig:ffr_evolution}. We will show in the following sections that the vertical mass flux is most likely responsible for this.

It is important to note that this process has to take place close enough to the planet. Otherwise, one can always find streamlines that satisfy $\zeta>1$, as through the keplerian shear in the CSD $\rho u^2$ increases more than linearly. In fact most of the streamlines far away from the planet at the gap edge fulfil this criterion. 

Hence, it seems that the opening of a free-fall region is permitted by the closeness of the $\zeta>1$ streamlines to the planet. This closeness to the planet is a result of the temperature, which determines the transition $\zeta=1$. 
Prior to the opening process, temperatures in the soon-to-be free-fall region are as cold as 75 K, which is even colder than assumed in some isothermal simulations, e.g. the work by \cite{zhu2016} where the midplane temperature inside $r_{\rm H}$ is truncated to 100 K. 

A last comment is in place in order to clarify the nature of the free-fall region in relation to numerical solutions found in other works. One might expect a planet to form a relative vacuum between the disc and itself, as for example in \cite{bethune2019} for strongly contracting planets, or as static solution for a given mass at low constant opacity, as ours. However, as the above analysis reveals, the mechanism for opening a free-fall region appears to be a dynamic process in conjunction with the thermodynamics of the spiral arms, rather than the properties of low-dimensional static solutions.

\subsection{CPD Rotation and CPD spiral arms in the midplane}
\label{subsec:cpdrotation}

The flow which enters the CPD through the midplane triggers the midplane accretion shock at around 50\% of all midplane angles. However, with only 50 K above the 200 K CPD background temperature, this shock is fairly weak and presumably not a significant influence for the entropy evolution of the CPD.
After the accretion shock, the gas quickly encounters the CPD spiral arms. Those spiral arms torque the flow significantly, but in a way as to transfer radial momentum to angular momentum, and hence boost the flow to higher fractions of $v_{\rm \theta}/v_{K}$ than those it initially possesses. This can be observed in Fig. \ref{fig:rotiation_and_accretion} where, following the streamlines, a region of high Keplerian rotation is evident after the encounter with the CPD spiral arms. This behaviour of the spiral arms is distinctly different from that in \cite{zhu2016}, where the CPD spiral arms act to reduce the angular momentum of the flow. 

The influence of spiral arms on the mean flow can be understood qualitatively in terms of interpreting the spirals as oblique shocks. An oblique shocks is a shock that has an inclination with respect to the local flow. Those have been already studied for a long time in the hydrodynamics literature \citep[and references therein]{kevlahan1997} where it is made clear that there is a vorticity jump imposed on the local flow through the shock. The vorticity jump is in general given by the density jump and the shock curvature with respect to the local flow. We note that because the CPD spiral arm shock is very 'fluffy' we cannot analyse the shock in a more quantitative manner as we analyse the CSD spiral arms in the next section and hence a fit of the post-shock values fails for the CPD spirals. 

The inner CPD boundary merits some comment as well. Here, gas does not flow directly into the planet, although it orbits at sub-keplerian velocity. This is partially an effect of substantial pressure support. On the other hand gas that originates in the CPD still flows into the planet, although in a vertical manner, that shall be explored further just below in Sect. \ref{sec:results_cpd_spiral}.

\begin{figure*}

\hspace*{+0.0cm}
\begin{subfigure}{1.0\textwidth} 
\centering
\includegraphics[width=\textwidth]{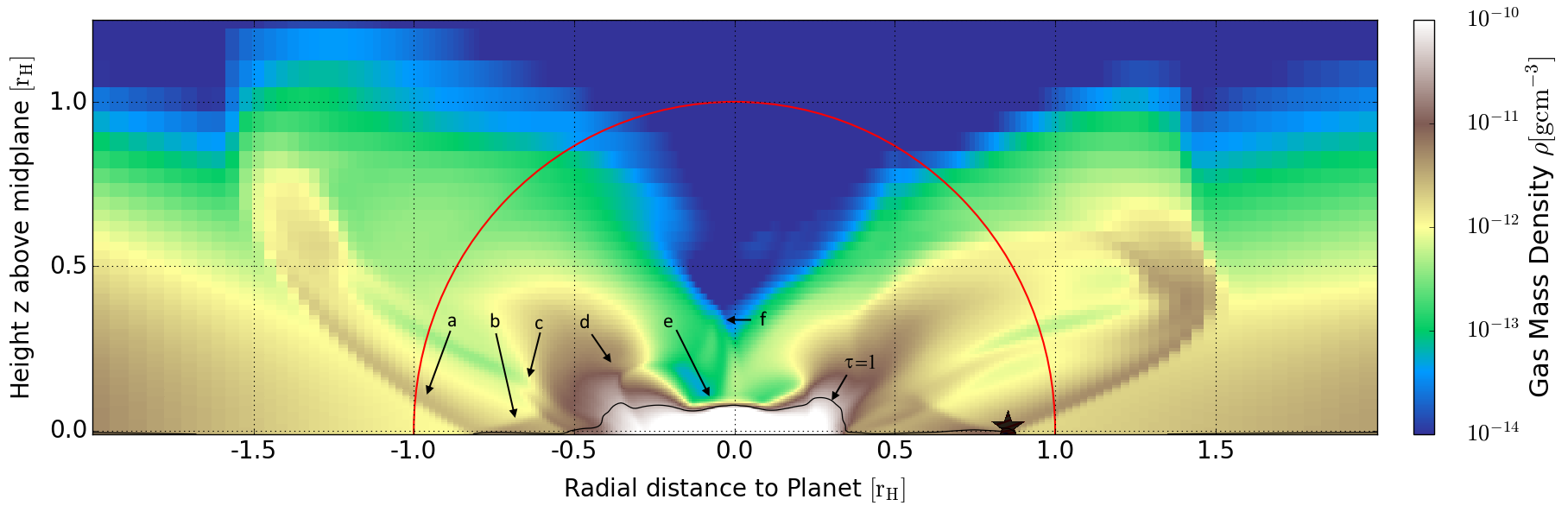}
\end{subfigure}%
\vspace*{+0.20cm}
\caption{Overview of density structures in a vertical cut through the non-circled star-symbol along the $y$$=$$0$-axis in Fig. \ref{fig:c1_midplane_dens_temp} for the nominal simulation run C1. The red circle marks the approximate Hill sphere. An approximate boundary of the CPD is the vertically integrated optical thin-thick transition, marked as $\tau=1$.
Features are labeled identically to the temperature plot (\textit{below}) with letters as follows: classical spiral arm (a), midplane free-fall region (b),  (midplane) accretion shock into the CPD (c), CPD spiral arm (d), vertical accretion shock (e), accretion funnel from colliding flows (f). }
\label{fig:c1_vertical_dens}

\begin{subfigure}{0.975\textwidth} 
\centering
\includegraphics[width=\textwidth]{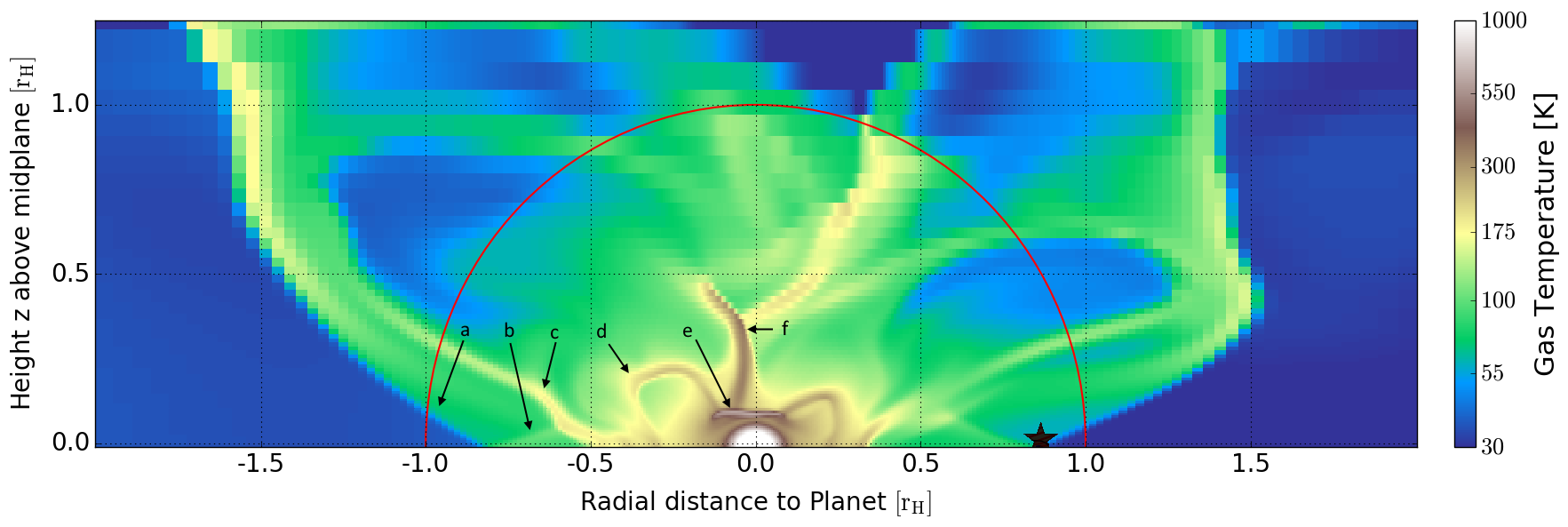}
\end{subfigure}%
\vspace*{+0.20cm}
\caption{Overview of temperature structures, for the same plane as in Fig. \ref{fig:c1_vertical_dens}. Some features are more clearly distinguishable than in the density plot (\textit{above}) in particular: the midplane accretion shock (c) and the vertical extension of the CPD spiral arms (d). Those features exhibit a radial asymmetry because the direction of the cut along the $y$$=$$0$-axis does not coincide with the symmetry axis of the CPD.}
\label{fig:c1_vertical_temperature}

\begin{subfigure}{0.975\textwidth} 
\centering
\includegraphics[width=\textwidth]{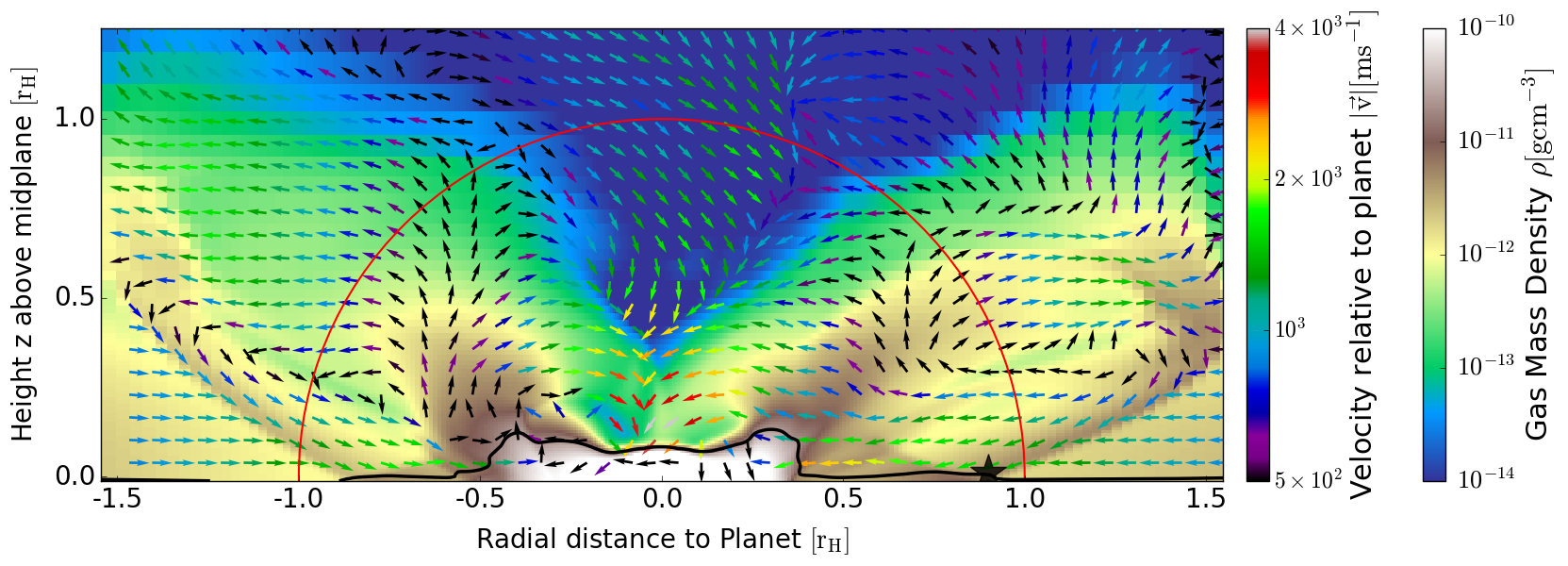}
\end{subfigure}%
\vspace*{+0.20cm}
\caption{Overview of a slice of the gas velocity field, in the same plane as in Figs. \ref{fig:c1_vertical_dens} and \ref{fig:c1_vertical_temperature}, arrows correspond to the radial and vertical components of the velocity on the slice and the colour shows the norm of the full 3D velocity. Most of the gas accreted by the planet enters the Hill sphere through the midplane, and through the tilted spiral arm shocks. The spiral arms give vertical kicks to the passing flows, forcing them to flow along the shock downwards into the midplane, where a region of increased compression is generated, marked with (b) in Figs. \ref{fig:c1_vertical_dens} and \ref{fig:c1_vertical_temperature}. This gas then decompresses and free-falls onto the CPD. \newline The majority of the incoming gas, however, is accreted by the planet eventually. This happens either after some residing time in the CPD or by directly flowing over the CPD and colliding with flow from the opposite side of the Hill radius. We note that the global meridional circulation noted in the isothermal runs of \cite{morbidelli2014} is an azimuthally averaged feature and hence not visible in this slice.}
	\label{fig:c1_vertical_flows}

\end{figure*}

\subsection{Vertical structure}
\label{sec:results_vertical_structure}

The vertical structure in density and temperature is plotted in Figs. \ref{fig:c1_vertical_dens} and \ref{fig:c1_vertical_temperature}, which we discuss in this subsection. There, vertical extensions of phenomena in the midplane can be identified. 
We additionally plot the vertically integrated optically thin-thick transition, marked with '$\tau=1$', as it delineates approximately the boundary of the CPD.

Feature labels in those figures are identical to those used previously in Figs. \ref{fig:c1_midplane_dens_temp} and \ref{fig:c1_midplane_profiles}, and are additionally described with $(r,z)$ coordinates in the text for clearer identification. Figs. \ref{fig:c1_vertical_dens} and \ref{fig:c1_vertical_temperature} have a radial extent chosen such that the density plot makes the gap density gradient evident, and their vertical extent is chosen such that the tilted structure of the spiral arms at our chosen lower density cutoff are still visible.

The CSD spiral arms are tilted in the vertical direction (feature (a)). As we will show later, this is a consequence of the disc thermodynamics-related process discussed earlier for the FFR. 
In the midplane, directly inward from the spiral arms, the FFR gas falls (feature (b)) onto the CPD. The free-falling gas from the CSD in the midplane can be best identified in the density plot, where it emanates from $(x,z)=(\pm0.9,0)$ at the position of the spiral arm shocks. From those positions it spreads out into a vertical fan, until it hits the CPD between $(x,z)=(\pm0.4,0)$ and $(x,z)=(\pm0.6,0.2)$. At those positions, the vertical extent of the CPD accretion shock is visible, as temperature maximum that traces a concave path (feature (c)).

\begin{figure}   

	\centering
	\includegraphics[width=0.45\textwidth]{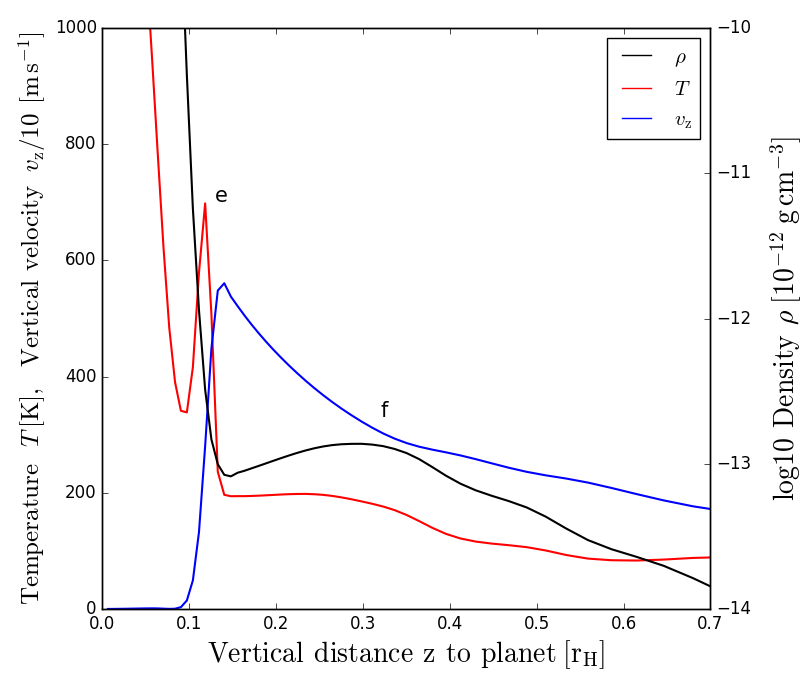}
         \label{fig:c1_vertical_profiles}

\caption{Profiles of temperature, vertical velocity and density as a function of height over the mid-plane. The letter labels correspond to those used in Fig. \ref{fig:c1_midplane_dens_temp} and Figs. \ref{fig:c1_vertical_dens} - \ref{fig:c1_vertical_flows}. At a height of up to $\tilde{z}=0.5$ above the planet, we find colliding streams (see also fig. \ref{fig:c1_vertical_flows}) that originate from overshooting the CPD after originally entering the Hill sphere through the midplane.
The resulting compression leads to a slight density bump (f), which is then accreted onto the planet and causes the vertical accretion shock (e).
%
}	\label{fig:c1_vertical_profiles}

\end{figure}

Next we find the CPD spiral arms and their vertical extent (feature (d)). The CPD spiral arms show the best contrast in the temperature data, and emanate from $(x,z)=(\pm0.3,0)$ towards higher altitudes and then towards the vertical accretion shock in a looped structure. We note that this loop-structure (particularly well seen in Fig. \ref{fig:c1_vertical_flows}) does not result from any vertical compression, as the gas flows parallel with respect to the loop. We also observe that this structure shows time-variance identical to the tidal arms in the midplane. From this we propose that the loop is in fact a vertical extension of the tidal arms, which must be driven by tidal resonances.

The most prominent vertical structure directly above the planet is the vertical flattening in density at $(x,z)=(0,z)$ that coincides with vertical free-falling gas and the vertical accretion shock seen at $(x,z)=(0,0.2)$ in the temperature (feature (e) in Figs. \ref{fig:c1_vertical_dens} and \ref{fig:c1_vertical_temperature}). This feature is accompanied by a vertical column of high-temperature gas, extending up to $(x,z)=(0,0.5)$ (feature (f) in Figs. \ref{fig:c1_vertical_dens} and \ref{fig:c1_vertical_temperature}). The high temperature column is a consequence of colliding flows from opposite sides of the planet. This flow collision cancels lateral velocity components and leaves the vertical velocity component nonzero, upon which the gas free-falls vertically into the planet.
This accretion flow is qualitatively similar to what has been postulated and numerically observed in the framework of Bondi accretion \citep{edgar2004}.

In order to have a more quantitative estimate of the effects of those features on local variables, we show the most important variables in vertical 1-D profiles in Fig. \ref{fig:c1_vertical_profiles} directly above the planet. There, it is evident that the colliding flows create a negative density gradient $\partial \rho/\partial z$ at $\tilde{z}>0.35$, but it free-falls towards the planet below this level, reversing this gradient. We now turn to describe the gas flow that interacts directly with the CPD.

\begin{figure*}
\hspace*{+0.5cm}	
  \begin{subfigure}{0.50\textwidth} 
   \centering
   \includegraphics[width=\textwidth]{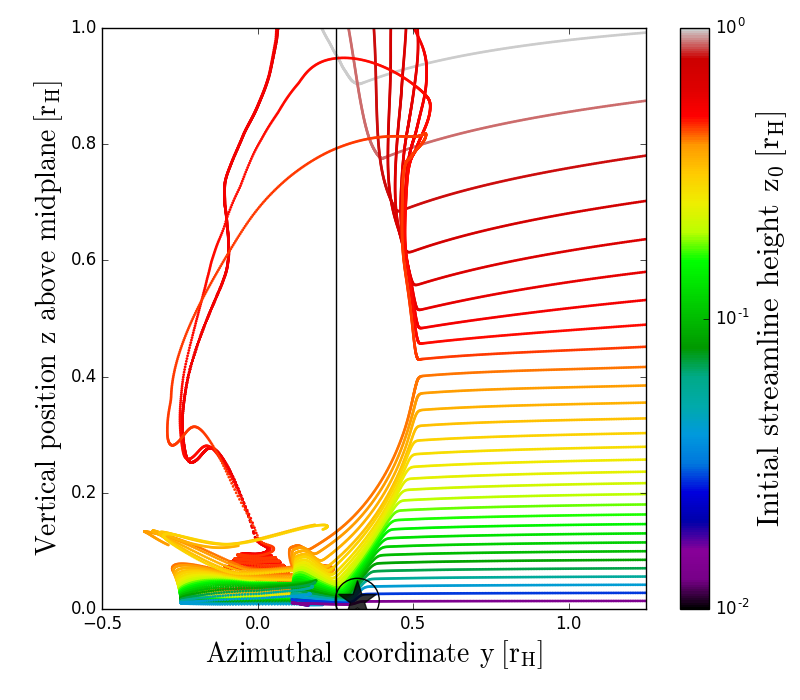}

	\end{subfigure}%
	\hspace*{+0.50cm}
	\begin{subfigure}{0.45\textwidth} 
   \centering
   \includegraphics[width=\textwidth]{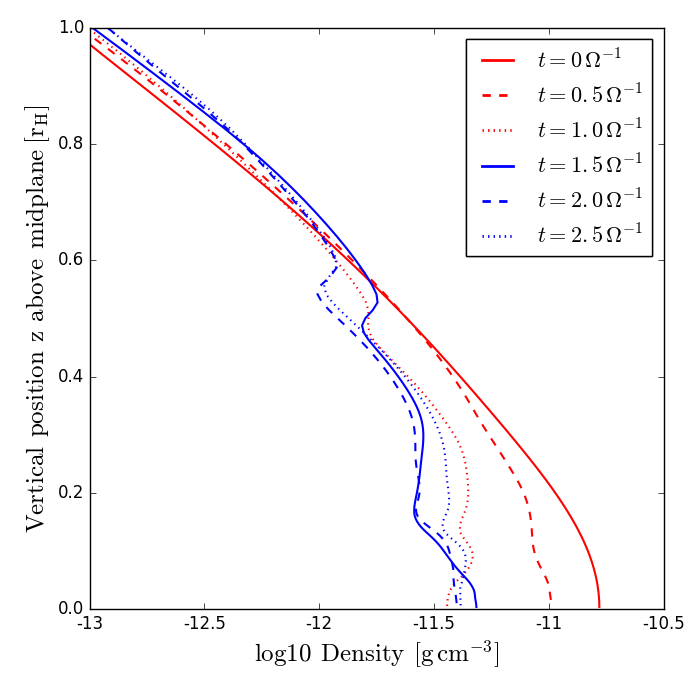}
	\end{subfigure} 
	      \caption{Vertical side-view of 3D-integrated streamlines (\textit{Left}) going through the circled-star-position in Fig. \ref{fig:c1_midplane_dens_temp}. Note that here we show the y-z plane. Streamlines are coloured according to starting height for purposes of distinguishing them. The circled star (bottom of the plot) is at the same 3-D coordinates as before. At around $z=0.41 r_{\rm H}$ a flow separation occurs, dividing the streamlines into strongly shocked upwards and strongly shocked downwards flows. Scanning through horizontally adjacent streamline families reveals that this phenonemon is responsible for leaving the entire column volume behind the shock devoid of streamlines with the ability to replenish the missing mass.
	      Evolution of the free-fall region at the vertical black line (\textit{Right}). Density profiles are shown evolving in time at the position indicated by the line in the left figure. The formation timescale of the FFR can be read-off as $\approx 2$$\Omega^{-1}$. For a density map during the formation process, see also Fig. \ref{fig:ffr_evolution}. }
\label{fig:shock_freefall_vacuum}

\vspace*{0.2cm}

	
	\hspace*{+0.50cm}
	\begin{subfigure}{0.45\textwidth} 
   \centering
   \includegraphics[width=\textwidth]{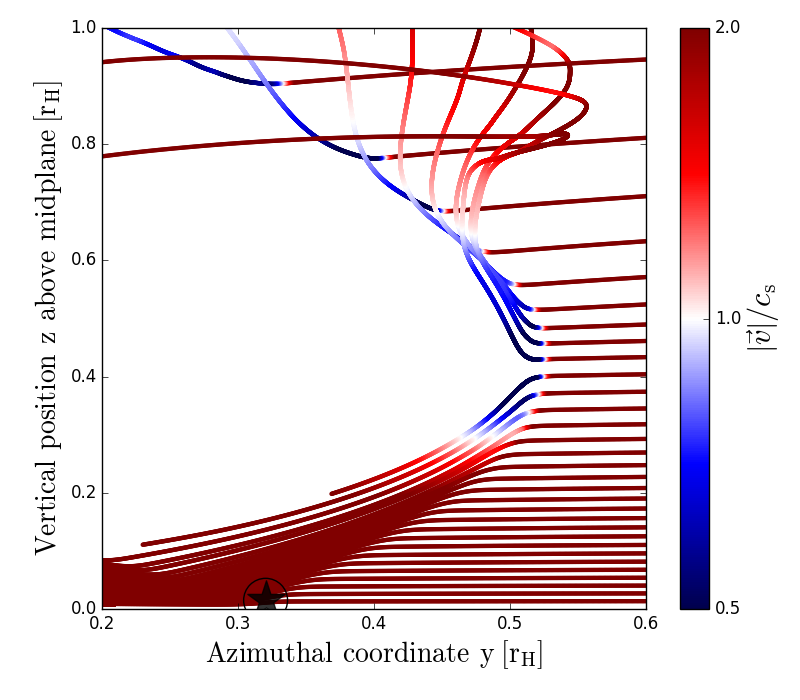}
	\end{subfigure}%
	\hspace*{+0.50cm}
	\begin{subfigure}{0.45\textwidth} 
   \centering
   \includegraphics[width=\textwidth]{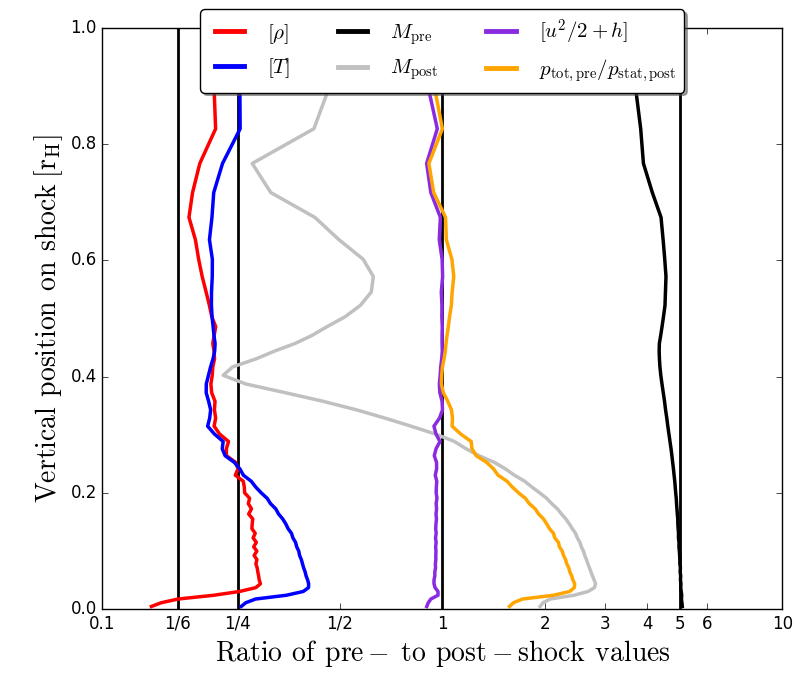}
	\end{subfigure} 
      \caption{Streamlines at the circled star from Fig. \ref{fig:c1_midplane_dens_temp}, with zoom-in on the shock and super-subsonic-transitions (\textit{Left}). The same streamlines are shown as in Fig. \ref{fig:shock_freefall_vacuum}, just with a limited range of azimuthal distances, for clarity. Note the asymmetry in the vertical, giving the shock an overall concave shape. Ratios of pre- to post- shock variables are shown for the same vertical position on the shock (\textit{Right}). 
      The shock jump values are important for explaining both the shape of the shock and the occurrence of the free-fall region.
      In the upper parts of the shock the ram-pressure of the incoming flow is insufficient to overwhelm the shock static pressure of the spiral arm, and the flow is forced to become subsonic (seen where $M_{\rm post}< 1$). Just below the flow separation at $z_{\rm sep}=0.41 r_{\rm H}$, as the density increases, the flow ram pressure becomes sufficient to overcome the static shock pressure. This happens below $z=0.38 r_{\rm H}$. In this region, the pre-shock supersonic flow also remains supersonic post-shock, albeit slowed down. The flow then continues to advect material from the post-shock volume, leading to the formation of the free-fall region. The region below $z=0.03 r_{\rm H}$ encounters in its first post-shock cell material coming from above, along the shock curvature. Those colliding flows lead to compressional heating at the 'foot' of the shock, increasing the density and temperature contrast. This foot region is important as it delivers mass to the midplane, keeping the midplane density jump alive.
      Along the entire shock, the adabatic ratio $[u^2/2+h]=1$ holds, revealing that radiative effects play only a minor role for the structure of this shock.
      }
         \label{fig:shock_mach_pressure}
        
      \vspace*{+0.20cm}   
\end{figure*}

\subsection{Vertical mass delivery to the CPD I: The free-fall region as an effect of vertically tilted spiral arms}
\label{sec:vertical_cpd_I}

In order to understand the details of the free-fall occuring in the midplane we have to develop an understanding of the behaviour of the mass column directly above it. We have seen in Sec. \ref{sec:midplane_freefall} how the free-fall region is opened as a consequence of overpressure and supersonic erosion in the midplane. In this section we will discuss how the vertical tilt of the spiral arms leads to a cut-off of mass supply to the entire volume of the free-fall region.

Two- or three-dimensional flows that pass through shocks tilted at an angle with respect to their flow direction behave like oblique shocks, i.e the flow component perpendicular to the shock surface is conserved, while that parallel to it experiences a discontinuity. As mentioned before, this is often expressed as vorticity jump, and this vorticity jump depends on the shock obliquity. Particularly in the physics of gap formation in protoplanetary discs the vorticity jump across the spiral arm shock in the midplane of protoplanetary discs \citep[and references therein]{li2005} is important. This is because one can understand the process of gap formation as the effect of the vorticity jump provided by the spiral arm flow patterns. 

We find an interesting analogy of this gap opening in the vertical direction.
The gas flowing from the CSD is initially stably stratified in the vertical direction. When encountering the tilted shock surface, this gas experiences a vorticity jump that redirects it. The direction into which the gas is redirected depends on the attack angle of the flow towards the shock. This becomes evident from inspecting the results from the 3D-integration of a selected family of streamline trajectories in Fig. \ref{fig:shock_freefall_vacuum}. Those streamlines pass through the spiral arm shock, at the position that was marked with a circled star in earlier plots.

There, the streamlines that encounter an upwards tilted shock surface are also tilted upwards, and vice versa.
Post-shock, the downwards flowing streamlines are forced to converge through redirection towards the midplane. This causes a region of compression along the flow of streamlines, but most of the streamlines are now directed away from the volume that is geometrically directly behind the spiral arm. We mark this volume simplistically with a black line.

Once the gas reaches the midplane, a region of high compression is created, which is the 'foot' of the shock. At this point, the gas is still $\sim$$0.9\,r_{\rm H}$ away from the planet, but can now fall onto the CPD through the free-fall region.

This vertical flow pattern is repeated for all streamline families that are horizontally adjacent. Hence, the entire columnar volume that forms the free-fall region cannot be replenished with gas by CSD gas due to the vertical tilt of the spiral arm shock (c.f. Fig. \ref{fig:shock_freefall_vacuum} \textit{right}). This explains how the state of lower density of the free-fall region is maintained, but not what the preconditions are for the vertical tilt of the spiral arm. We will investigate this now.

\subsection{Vertical mass delivery to the CPD II: Vertically tilted spiral arms as an effect of competition between ram and static pressure, initiating FFR formation }
\label{sec:vertical_cpd_II}

In order to learn further about the physical conditions behind the spiral arm shock and shed light on the occurence of the free-fall region, it makes sense to investigate the pre-shock conditions. 

For this purpose, we take a more detailed look at the same family of streamlines as in the last subsection. We zoom in radially onto the shock, and plot where the transition from locally super- to subsonic gas flow occurs. Because the shock evolves only very slowly for the first two orbits and then reaches a steady state, the co-rotating frame of the planet and the shock-rest frame coincide. Hence, we analyse the shock data in the co-rotating frame of the planet. 
The result can be seen in Fig. \ref{fig:shock_mach_pressure} (\textit{Left}). In this figure, the shock surface can be traced by eye, by following the sharp increase/decrease in the vertical velocity component, which is at the points of sharp upwards/downwards turns of the gas. We will refer to this upward/downward transition as the \textit{vertical flow separation} at a height $z_{\rm sep}=0.41$.
A curious pattern emerges as function of the vertical direction on the streamlines. The post-shock flow changes its Mach-number from being sub-sonic to remaining super-sonic, just at about the height of $z_{\rm sep}$.

In order to analyse this behaviour further, we plot the pre- to post-shock jumps for some important shock quantities on the same streamlines, along the shock surface in Fig. \ref{fig:shock_mach_pressure} (\textit{Right}). First, the jumps of density and temperature are of interest. 
The bracket $[x]$ denotes the jump ratio $[x]=x_{\rm pre}/x_{\rm post}$.
Some physically important values for our parameters are marked in the plot.

In an adiabatic shock with our adiabatic coefficient of $\gamma=1.4$, the jumps for $[\rho]$ and $[T]$ should approach $[\rho]= 1/4.5$ for $M_{\rm pre}=4$ (with $[\rho]\rightarrow 1/6$ for infinitely strong shocks) and $[T] \approx 1/4$ for $M_{\rm pre }=4$ (with $[T] \rightarrow \infty$ for infinitely strong shocks). From the data it is evident that the behaviour of the spiral arm in the upper layers, for $\tilde{z}>z_{\rm sep}$, is consistent with that of adiabatic shocks.
 
For $\tilde{z}<z_{\rm sep}$ the shock behaviour changes. An important possbility of modifying the shock jump-values is the radiation of the shocked gas. Both sides of the shock radiate energy, and as $[T]$ is unbound, important asymmetries in radiative fluxes can arise. \cite{mihalasmihalas} (ch. 104) show how the Rankine-Hugoniot jumps for a radiating shock will change compared to the adiabatic case:
in a radiating shock, the temperature jump is expected to be more moderate compared to the adiabatic case, because both sides of the shock radiate back at each other, while the density jump should be increased.

\begin{table*}
\label{tab:temperature_data}
\centering
\caption{Phenomenology of spiral arms, occurrence of free-fall regions, free-fall region evacuation, repeated analysis along the same initial streamline family as for run C1. }
\begin{tabular}{c c c c c c c}        
\hline\hline                 
Run + label & Spiral arm shape & Free-fall onto CPD? & FFR evacuated? & $T_{\rm pre}$ & $T_{\rm post}$ & $P_{\rm tot,pre}/P_{\rm stat, post}$\\    
\hline
C1 "nominal" & concave & Yes & Yes & 30 & 90 & >2 \\
C2 "belllin" & concave & Yes & Yes & 33 & 120 & >2\\
H1 "high opacity" & convex & No & No & 70  & 90 & 1\\
H2 "deep" & straight & No & No & 105 & 170 & 1\\
\label{tab:temperature_data2}
	\end{tabular}
	\\[5pt]
	\caption*{(*) Sample values pre-shock, in order to showcase significant differences. Values are taken above the midplane, to avoid effects of a 'foot'-type post-shock region. The pre-shock temperature for the 'cold' simulations, C1 and C2 are essentially the disc temperatures. For the hotter simulations H2 and H1, the pre-shock temperatures rise significantly due to radiative precursors.}
\end{table*}

The jump-data we see in $[\rho]$ and $[T]$ for $\tilde{z}<z_{\rm sep}$ is therefore inconsistent with both adiabatic and radiative shocks, at least using the co-rotating planetary frame as shock-rest frame. Hence, although we see discontinuities in the data, we need to search for another possibility in order to explain this data.

Instead, we show that the ratio of pressures $\zeta$ (as defined in Eq. \ref{eq:pressure_ratios}) explains the data well. Intuitively, it is clear that for $\zeta>1$ the post-shock gas can remain supersonic. Pushing past the shock should be only possible for incoming gas, if the otherwise unsurmountable pressure-barrier of the shock is instead felt as a relatively insignificant 'bump on the road'. This is an explanation that seems to agree well with our previous analysis from Sec. \ref{sec:midplane_freefall} in relation to Fig. \ref{fig:ffr_evolution} and now in the vertical data in Fig. \ref{fig:shock_mach_pressure}.  

The advection of post-shock material leaves the question of why the shock does not disappear altogether and appears static. The only explanation we could find for this behaviour was the vertical delivery of mass towards the 'foot' of the shock, as evidenced by the additional compressional heating for $z<0.03\,r_{\rm H}$ and shown in Fig. \ref{fig:shock_mach_pressure}. The vertical mass fluxes per streamline are about a factor $\sim$30 lower than those advected through the shock in the midplane, but this is the only source of mass available to keep the midplane shock in existence. 

As a sanity check in order to spot any anomalies we plot the total energy jump $[u^2/2+e+P/\rho]=[u^2/2+h]$, where $h$ is the enthalpy, and this yields a constant value of $1$ across the whole shock. This confirms that the shock obeys overall correct thermodynamics, and the structures we see do not originate in numerical effects.

\begin{figure*}
\hspace*{+0.5cm}	
   \centering
   \includegraphics[width=0.7\textwidth]{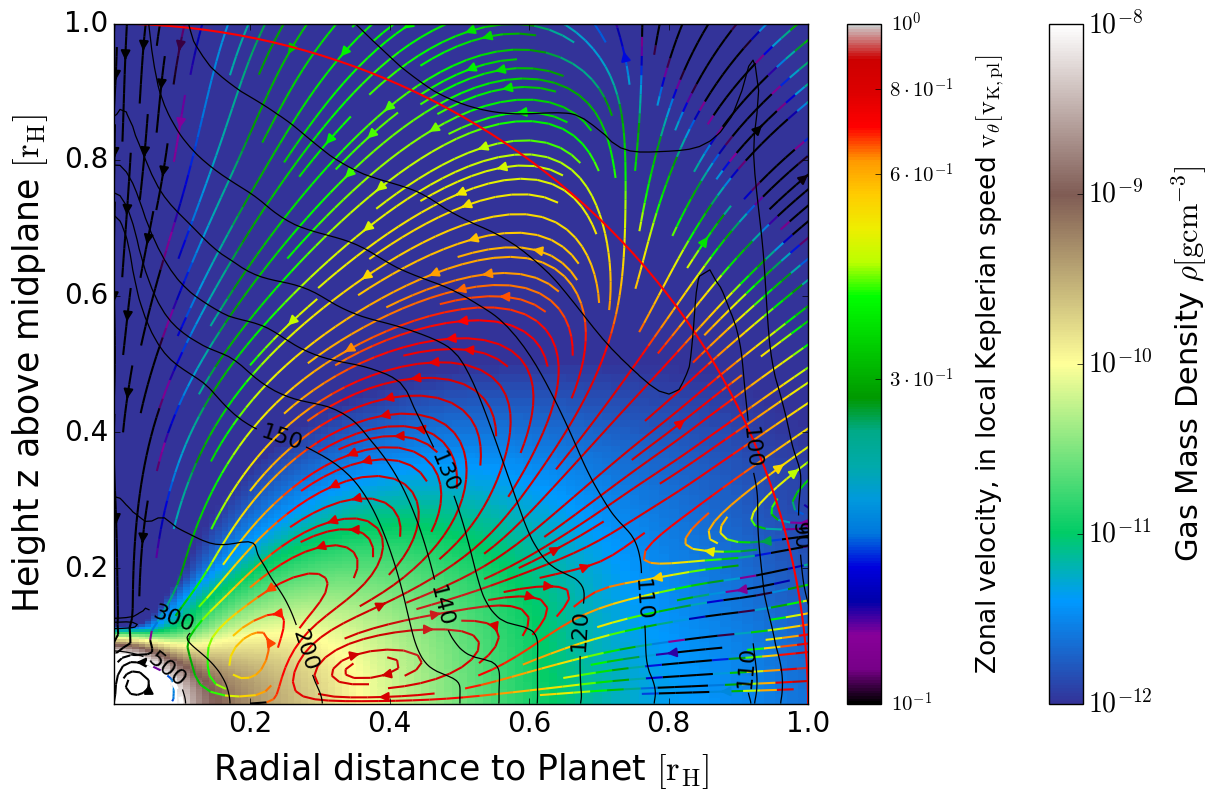}
\caption{Cylindrically averaged vertical structure of flows, density and temperature in K, shown on the contours shown against planetocentric radius and height. Deviations from a spherically symmetric temperature profile are due to compressional heating for the free-falling and CPD-overshooting midplane flows, and due to efficient radiative cooling in the vertical. Density colours show the toroidal structure of the CPD. Streamline colours indicate the fraction of keplerian rotation on each flow line, helping to estimate gas residence time on each particular orbit. Streamlines of high keplerian rotation will therefore orbit more perpendicular to the plot than follow the streamlines, and vice versa.}
	\label{fig:c1_flows_cylindrical}
\vspace*{+0.20cm}

\end{figure*}

\subsection{Vertical mass delivery to the CPD III: Spiral arm tilt as effect of the ram pressure}
\label{sec:vertical_cpd_III}

The dynamic overpressure in the lower parts of the shock, for $\tilde{z}<z_{\rm sep}$, helps explain the tilted vertical structure of the spiral arm structure. The entire region of overpressure is being pushed towards the planet, which distorts the shape of the spiral arm.

We also note that in our other simulation runs, we find agreement with this principle. A convex shape of the spiral arms seems to be a predictor of the existence of a free-fall region, and we now know that the convex shape is tied to the thermodynamics and the intrinsic cooling capability of the envelope. 
We indicate the qualitative features of the other shocks in Tab. \ref{tab:temperature_data2} for simulations C1-C2. Simulations with a higher opacity or a deeper potential generally do not display a FFR, due to their increase in pressure support of that region.

\subsection{Planetary accretion through the CPD, the 3D circulation inside the CPD and its inner truncation radius }
\label{sec:results_cpd_spiral}

After discussing the mass delivery we now focus on describing the state of the gas that is entering the CPD from the inner CPD radius, before it starts to orbit in the CPD. 
It was already noted in Sec. \ref{subsec:cpdrotation} and Fig. \ref{fig:rotiation_and_accretion} that gas enters mainly through the midplane and is bumped to high values of $v_{\theta}/v_{\rm K}$ through the action of the CPD spiral arms. This mechanism has more consequences in the vertical, which we shall explore in this subsection. We already showed the general flow structure as 2D-cut in Fig. \ref{fig:c1_vertical_flows}, but for a full understanding of the flows it is necessary to investigate the average circulation inside the Hill-sphere, which we show in Fig. \ref{fig:c1_flows_cylindrical}. 

Once passed through the CSD spiral arms, the mass flux enters the Hill sphere predominantly through the midplane, as previously seen in Fig. \ref{fig:shock_freefall_vacuum} (\textit{Left}). 
Mass fluxes from higher altitudes decrease rapidly by orders of magnitude in strength, because the density decreases in altitude and the midplane temperature is low. This results in an integrated per-streamline-flux of $0.1 \%$ in the vertical direction, compared to the horizontal.

The free-falling gas coming from the CSD and the spiral arms enters the CPD after being slowed down in the midplane accretion shock, which shows a concave shape, the opposite of the spiral arms. This can be seen in Fig. \ref{fig:c1_flows_cylindrical}, together with the associated density and temperature structures of the CPD. There, vertical kinks in the temperature contours indicate changes in the compressional heating of flows, while the density structure of the CPD is relatively simple, at a high aspect ratio of $H/r\approx 0.1-0.2$. Streamlines that pass the CPD boundary at around $r\approx 0.4-0.5 r_{\rm H}$, exit again in the vertical direction, but further inward, starting at $(\tilde{r},\tilde{z})\approx (0.25,0)$. This exit is a comparatively slow process, as the gas is redirected on circular orbits with high $v_{\theta}/v_{K}$ after the CPD accretion shock and only slowly spirals up- and outwards from the planet for $\sim$$10-20$ CPD orbits. 
This process terminates at the height where the CPD spirals end (previously feature (d) in Fig. \ref{fig:c1_vertical_temperature} and here at $(\tilde{r},\tilde{z})\approx (0.4,0.3)$). This is indicative that this slow up/outwards migration of streamlines seems to be related to the kicks by the CPD spirals at every orbit, similar to the boosting that has been seen in the midplane (compare this to Fig. \ref{fig:rotiation_and_accretion}), just with a vertical component added to it. After rising to the height of CSD spiral termination, the gas moves towards the planet much faster.

\begin{figure*}
   
  \hspace*{+0.0cm}
  \begin{subfigure}{0.44\textwidth} 
   \centering
   \includegraphics[width=\textwidth]{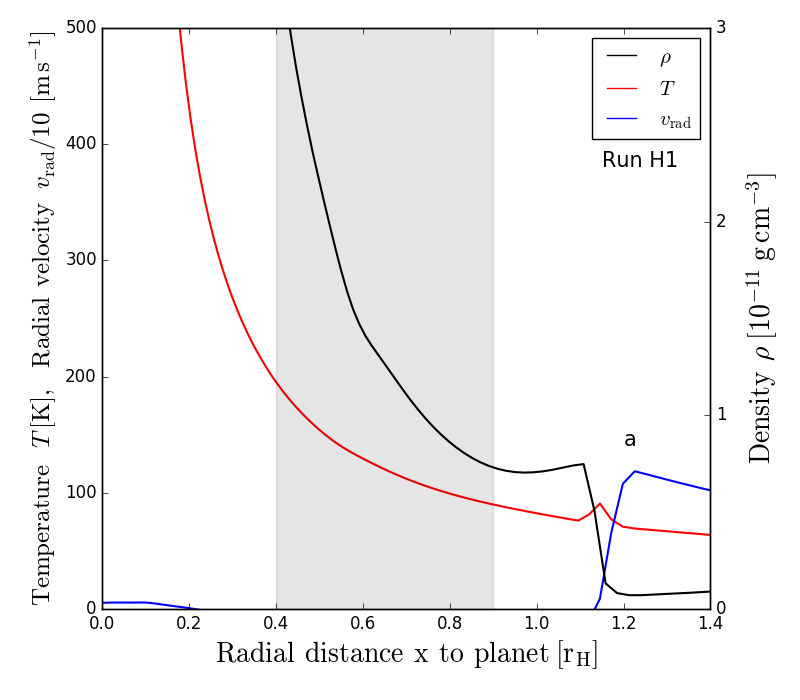}
	\end{subfigure}%
	\hspace*{+0.50cm}
	\begin{subfigure}{0.44\textwidth} 
   \centering
   \includegraphics[width=\textwidth]{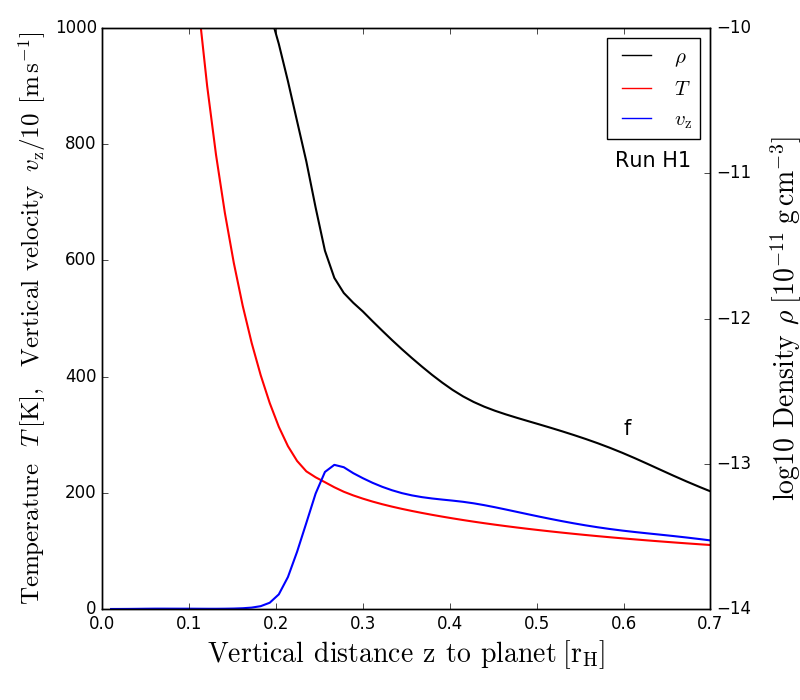}
	\end{subfigure} 
	
	 \hspace*{+0.0cm}
  \begin{subfigure}{0.44\textwidth} 
	\centering
	\includegraphics[width=\textwidth]{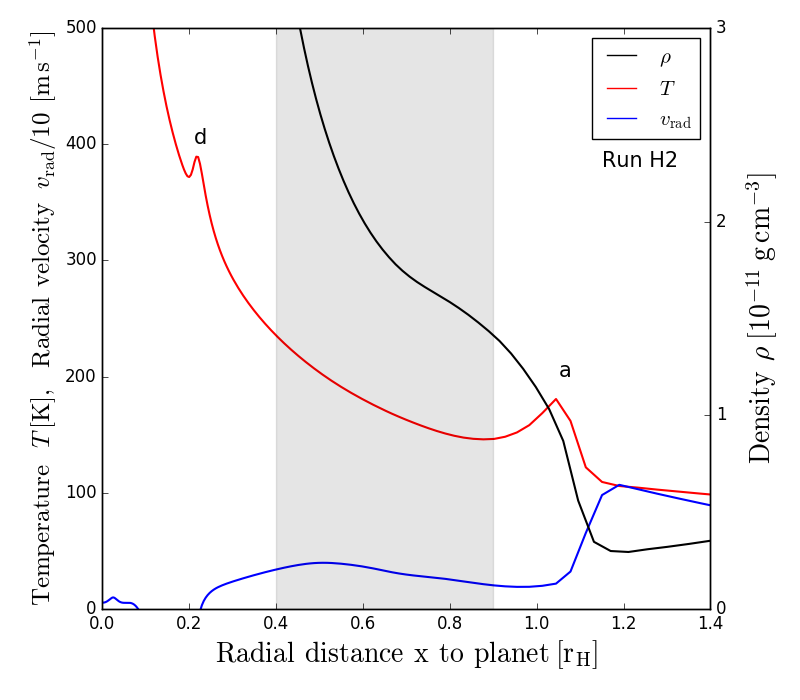}
   \end{subfigure}%
    \hspace*{+0.5cm}	
	  \begin{subfigure}{0.44\textwidth} 
	\centering
	\includegraphics[width=\textwidth]{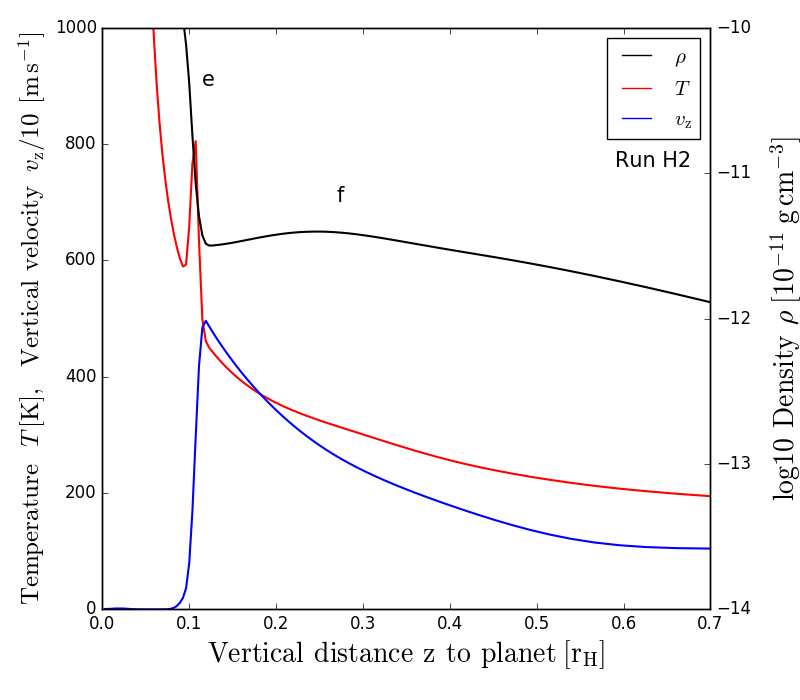}
   \end{subfigure} 
   
	 \hspace*{+0.0cm}
  \begin{subfigure}{0.44\textwidth} 
	\centering
	\includegraphics[width=\textwidth]{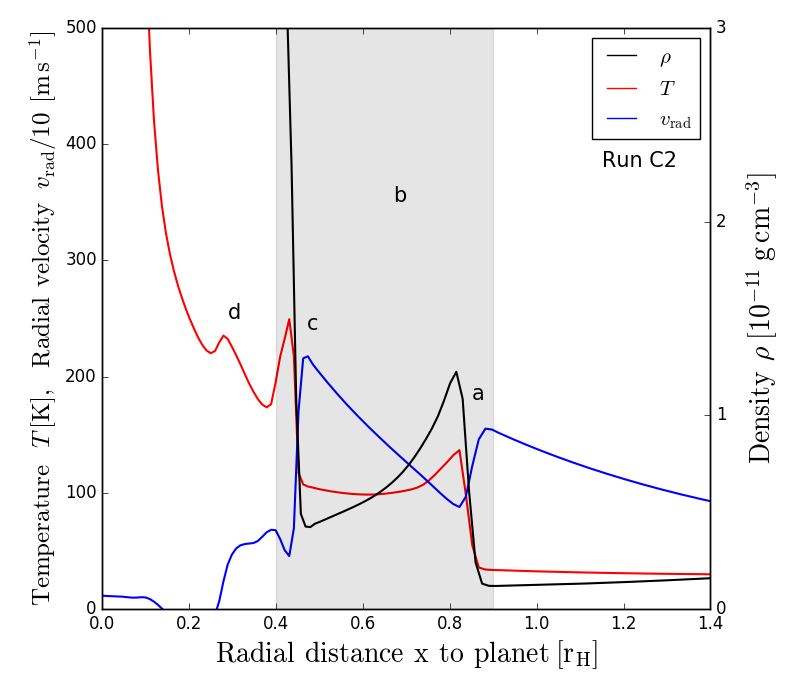}
   \end{subfigure}%
    \hspace*{+0.5cm}	
	  \begin{subfigure}{0.44\textwidth} 
	\centering
	\includegraphics[width=\textwidth]{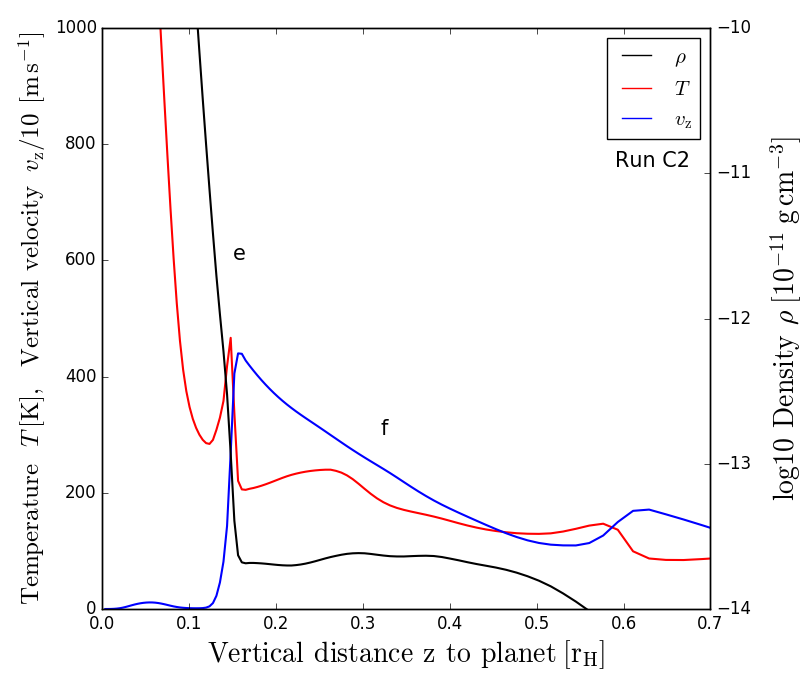}
   \end{subfigure} 

\vspace*{+0.20cm}
\caption{
Overview of slices through flow and structure variables, identical to Fig. \ref{fig:c1_midplane_profiles} and \ref{fig:c1_vertical_profiles}, only now for the runs H1 ("high opacity", \textit{top}), H2 ("deep", \textit{middle}) and C2 ("Bell \& Lin", \textit{bottom}) orbit 5 in the midplane (\textit{Left column}) and the vertical values (\textit{Right column}). The letter labels indicate the same physical features as before: CSD spiral arm shock (a), same volume as the free-fall region in run C1 for comparison (b), shock due to free-fall of gas from the spiral arms towards the CPD (c),  temperature bump due to the CPD spiral arms (d), vertical accretion shock (e) and accretion column from colliding flows (f). It becomes evident that the envelope with high opacity suppresses structural and flow features, consistent with the literature  \citep[e.g.][]{lambrechts2019}. A deeper potential raises temperatures even above those of the low-opacity run, but only suppresses the formation of a free-fall region. The Bell \& Lin opacities recover all features we found previously in the nominal run.}
	\label{fig:overview_opacities_profiles}
	
\end{figure*}

\begin{figure*}
\hspace*{-0.5cm}	
  \begin{subfigure}{0.47\textwidth} 
   \centering
   \includegraphics[width=\textwidth]{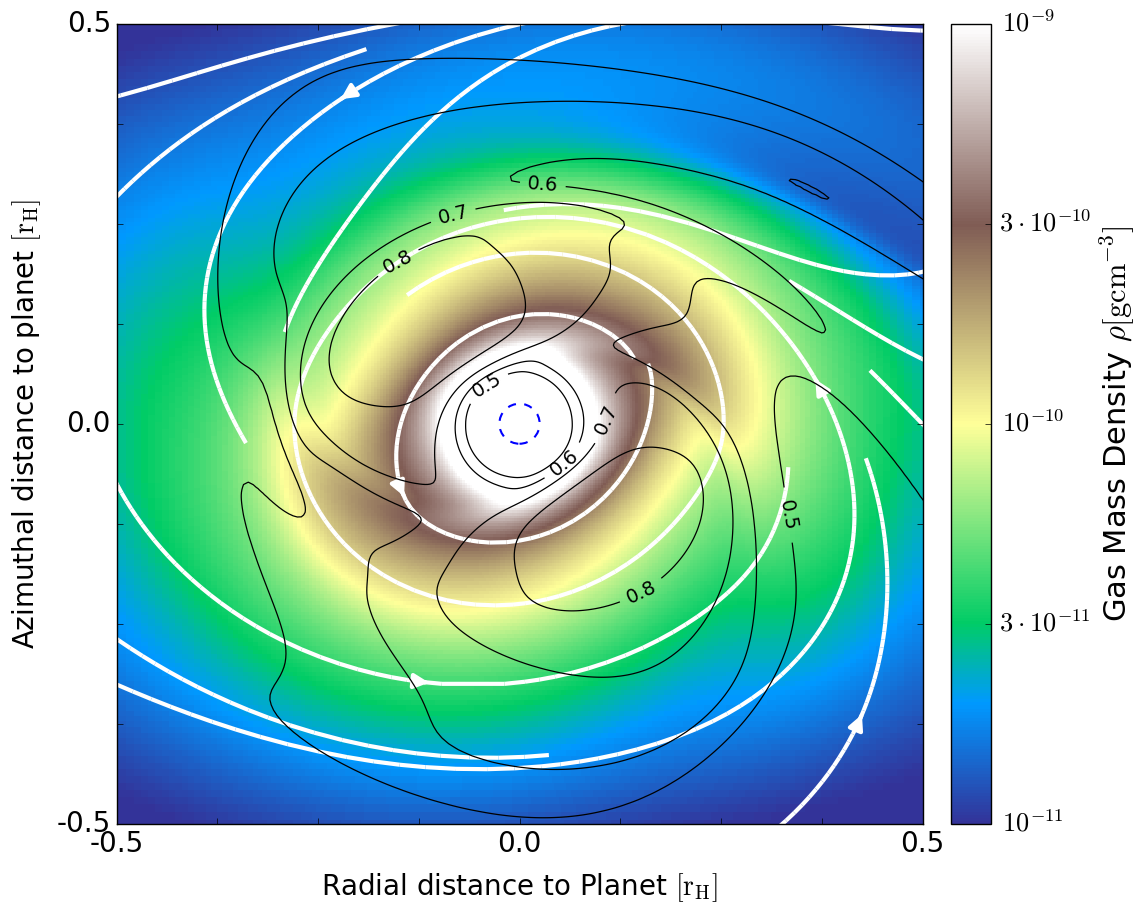}
	\end{subfigure}%
	\hspace*{+0.50cm}
	\begin{subfigure}{0.57\textwidth} 
   \centering
   \includegraphics[width=\textwidth]{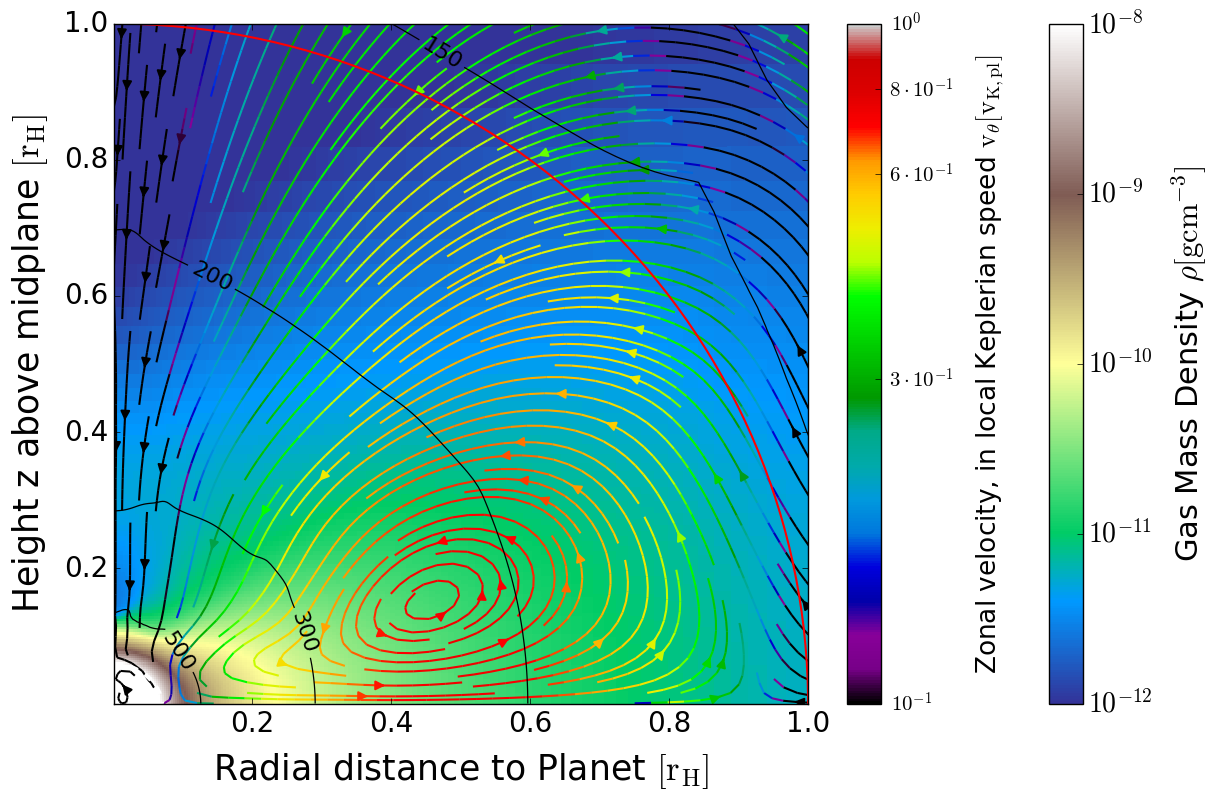}
	\end{subfigure} 
	\vspace*{0.2cm}
\caption{Midplane rotational structure for the run H2 (\textit{Left}), with $|\vec v|/v_{\rm Kep}$ as contours. Run H2 has a four times deeper potential compared to run C1, which is shown in Fig. \ref{fig:rotiation_and_accretion}. The gravitational smoothing length is marked with the blue, dashed circle. It becomes evident that the deeper potential does not extend the CPD necessarily much closer to the planet, rather the contrary. As pressure support increases, the maximum rotational support drops below 90\%. The physical extent of the strongly rotating gas remains similar to C1. Cylindrically averaged structures are shown with the temperature as contours (\textit{Right}). The observed flows are similar to the isothermal results presented in \cite{tanigawa2012} and consistent with previous high-opacity results from \cite{lambrechts2019}.}
	\label{fig:h1_all_flows}
\vspace*{+0.20cm}

\end{figure*}

This gives the gas enough time to cool, upon reaching higher into the CSD atmosphere. Viscosity lets the streamlines drift inwards slowly, until the streamlines reach the evacuated region directly above the planet and then fall towards it. Of those streamlines that reach the inner $\sim$ $0.2 r_{\rm H}$ a small part goes into a circulatory motion caught between $r_{\rm s}=0.1 r_{\rm H}$ and $0.3 r_{H}$, corresponding to the circulation seen earlier in the midplane in Fig. \ref{fig:rotiation_and_accretion}.

Mass accretion through the midplane therefore does not seem to deliver mass very efficiently to the CPD: the gas is merely passing through, albeit in a fairly complex manner, until it finally lands inside the smoothing length of the planet. Except for the small region between $r_{\rm s}=0.1 r_{\rm H}$ and $0.3 r_{H}$, which can compensate for lower keplerian rotation with radial-vertical circulation, we do not find any region in the CPD which would possibly permit long residence times for dust particles. This indicates that our simulation results pose a too early stage of the CPD life in order to form moons.

\section{Discussion and support simulations: Comparison with entire data set, CPD masses and planetary accretion rates}
\label{sec:results_comparison}

The previous section described the physical properties of the gas flow, density and temperature for the nominal simulation run. We now begin to introduce and compare the simulation runs for a number of important parameter variations. On one hand, for the discussion of CPD properties, we limit ourselves to comparing runs C2-H2 to C1, while on the other hand, in order to place our results into a larger context of growing protoplanets, we discuss mass accretion rates into all planetary envelopes and include runs m1-m5 into the discussion.

\subsection{Comparison to other simulation runs and robustness of C1 results}

In order to simplify the comparison of run C1 to our other Jupiter-mass simulation runs, we avoid plotting the full complexity of data shown previously. We only plot 1D-profiles in the midplane and the vertical as this is sufficient to comment on the structure of the envelopes of the simulations C2-H2 in Fig. \ref{fig:overview_opacities_profiles}.

Run H1, possesses a hundred times higher, constant opacity compared to C1. 
This parameter change is motivated by the possibility of dust-enriched, ISM-like material flowing into the feeding region of the giant planet, being distributed into the gas and hence increasing the ambient dust opacity. It also serves as an important comparison to earlier work using such high opacities \citep{judith2017,lambrechts2019}.
The physical extension of the envelopes in H1 is large enough as to preclude large free-fall distances, and additionally the temperature is everywhere high enough so that none of the infalling gas becomes supersonic. Hence, the paucity of physical features in H1 (see the top row of Fig. \ref{fig:overview_opacities_profiles}) is consistent with those earlier works, while the Hill-sphere is still accreting vigorously at a rate of $\dot{m}_{\rm H1} = 1.0\times 10^{-2}\, \rm m_{\oplus} \ \, yr^{-1}$. That those features are physical, rather than numerical, is supported by the fact that we ran H1 with the same numerical parameters as C1, the only difference being the opacity.

Run H2 possesses a four times deeper potential (and corresponding higher numerical resolution) compared to C1. This potential depth of $\tilde{r}_{\rm s}=0.025$ corresponds to a physical size of $r_{\rm s} = 12 R_{\rm Jup}$. This physical size, while being still too large for a young protoplanet, is an improvement over our nominally used potential depths and those of our previous works, but limits the simulation speed significantly.
The deeper potential causes strong initial compression of the infalling gas and hence high temperatures throughout the whole envelope. The central temperatures reach about 13000 K, which is very high, but this may be an artifact of our constant $\gamma$ treatment that ignores the dissociation of the hydrogen molecule. The larger extent of the planetary boundary condition inwards serves as a check for the robustness of our inner CPD boundary. 

As can be seen in Fig.\ref{fig:overview_opacities_profiles} (\textit{middle row}) the emergence of a free-fall region is suppressed when deepening the potential in H2. We find that this is consistent with our previous analysis of spiral shock shape and pressure ratio $\zeta$. The intense radiation field released by the deep potential is shining on the spiral arm and stabilizes it by increasing the static pressure. Data on the spiral arm phenomenology for runs C1, C2-H2 were already summarized in Tab. \ref{tab:temperature_data2}.

The rotational state of the H2 envelope is only weakly impacted by the elevated temperature. As can be seen in Fig. \ref{fig:h1_all_flows} (\textit{Left}), the envelope achieves above $80\%$ Keplerian rotation. Hence, contrary to \cite{judith2016} we obtain keplerian rotation values that are equally high as in their their cold simulation, but with the planetary temperatures of their hot simulation.
This is because the lower opacities compared to \cite{judith2016} allow for higher temperature gradients between deep planetary envelope, so that the planet can be very hot but still allow for the CPD to cool down.
It seems that the central planetary temperature does not determine whether a CPD or envelope forms, but rather how far inwards it extends.

The CPD also possesses a flow separation between inner and outer envelope at $r\approx 0.1 r_{\rm s}$, at the same radius as run C1. This demonstrates the robustness of this inner CPD edge against our numerical parameters. However, the global flow inside the Hill-sphere has changed significantly, as is evident from Fig. \ref{fig:h1_all_flows} (\textit{Right}). 

This global flow pattern in H2 follows a global circulation pattern, bearing similarity to that one seen in \cite{tanigawa2012} (obtained with locally isothermal simulations), and is similar enough to that one in H1, which is why we do not show the latter simulation separately. The reason for the similarity does not seem to be the fact that the H2 CPD is vertically quasi-isothermal, as the vertical temperature profile is similar in C1, but rather the radial structure of the pressure gradients.
Interestingly, the accretion rates do not differ that strongly when comparing C1 and H2. The accretion rate into the H2 envelope is
$\dot{m}_{\rm H2} = 1.8\times 10^{-2}\, m_{\rm \oplus} \ \, yr^{-1}$ whereas the accretion rate into the C1 envelope was $\dot{m}_{\rm C1} = 3.5\times 10^{-2}\, m_{\rm \oplus} \ \, yr^{-1}$.

Finally, run C2 uses the more realistic opacities of \cite{belllin1994}, instead of constant opacities. This set of opacities features two important opacity transitions, which correspond to the sublimation temperatures and densities of water and silicates. This run has a long burn-in time of the simulation because of the non-linear  time-evolution of the envelope opacities, but eventually finds a quasi-steady state. While the exact values of density and temperature in C2 differ slightly compared to C1, we recover all simulation features of C1 also with the non-constant Bell \& Lin opacities. This is an important data point showing the validity of our previous analysis.

We need to stress that the global flow patterns inside the Hill-spheres of our Jupiter-mass planets seem to divide along two different kinds: C1 and C2 both feature a free-fall region due to their low envelope temperatures and their flows are very similar and complex. H2 and H1 being more 'hot' simulations feature a simple global circulation accreting significant mass through the vertical direction, albeit with the same inner CPD cut-off as the 'cold' circulations.

For both types of circulation, the accretion rates remain high and this will be the subject of further discussion in the next subsection.
Most of the mass ($\approx$99\%) entering the Hill sphere enters through the midplane and is then either processed through the CPDs, after residence times of several tens of CPD orbits, or evades the CPD and falls into the planet at a vertical angle of $\approx$45$^{\circ}$ directly.

Those simulation results demonstrate that a low mean opacity is key for the development of circumplanetary discs. In star-forming environments this could be achieved by having a reduced total dust mass in the $\mu$m grains, which dominate cooling rates, if present. Another possibility to obtain lowered mean opacities is via growing the same dust mass to mm-sizes on average. Evidence for dust growth in prestellar cores \citep{chacon2019} has been reported, and even the presence of mm-sized grains has been inferred indirectly in protoplanetary discs \citep{harsano2018}. Therefore our choice of mean opacity values is well-supported by observational evidence.

\begin{figure*}	
	\hspace*{+0.50cm}	
	\begin{subfigure}{0.45\textwidth} 
   \centering
   \includegraphics[width=\textwidth]{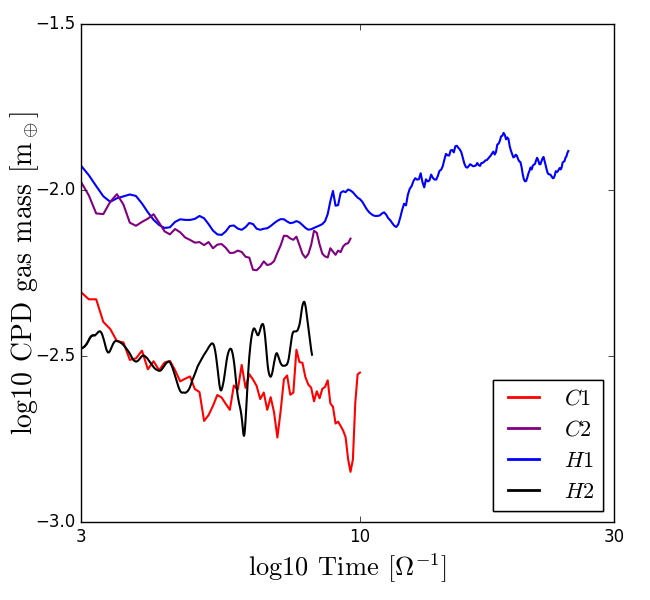}
	\end{subfigure} 
	\hspace*{+0.50cm}
    \begin{subfigure}{0.45\textwidth} 
   \centering
	\includegraphics[width=\textwidth]{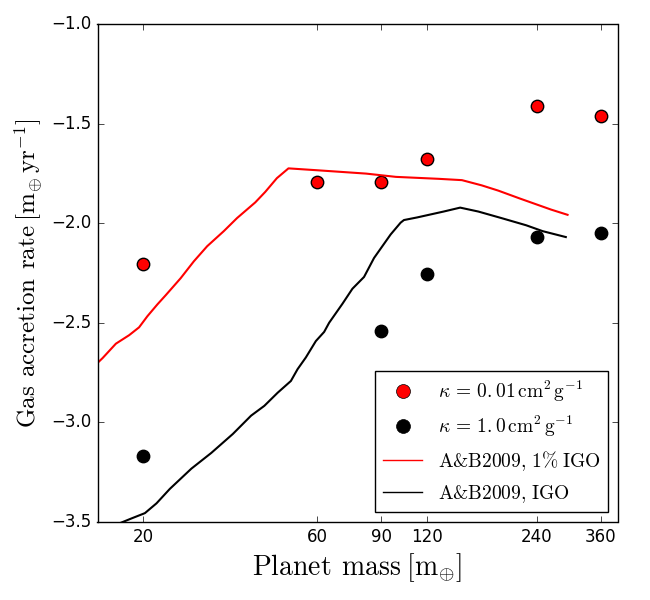}
	\end{subfigure}%
    
\vspace*{+0.20cm}
\caption{Circumplanetary disc masses for the Jupiter-mass planets (\textit{Left}) seem to remain stable after the initial 2 orbits of ramp-up time of $r_{\rm s}$. The CPDs, except for C1, start accreting with gas accretion rates in the range of $1-4\times10^{-5}\,\rm m_{\oplus} \, yr^{-1}$, corresponding to growth times for the Galilean satellites between 80 kyr and 0.32 Myr. Simulations are not run for the same number of orbits due runtime restrictions.
Mass accretion rates onto the planets as function of planet mass and opacity (\textit{Right}). Note similarities to \cite{lambrechts2019}, which used used Bell \& Lin opacities for a similar mass range. Earlier results from \cite{ayliffe2009a} using SPH simulations and interstellar grain opacities (IGO) agree with ours in the mid-mass range, our low masses accrete visibly more and the high masses do not seem to converge on a single disc-limited accretion rate. One could speculate that this is related to the strong disc inflow that triggers the CSD spiral arm tilt, effectively increasing the mass inflow.}
	\label{fig:accretion_and_cpdmasses}
\end{figure*}


\subsection{Gas accretion into CPD, CPD masses and radial mass profiles}
\label{sec:gas_accretion_cpd}

While the planetary accretion rates are vigorous for the Jupiter-mass planets, it is interesting to note that our CPDs do  accrete as well, which can be seen in Fig. \ref{fig:accretion_and_cpdmasses} where we plot the CPD masses, defined as the mass between the shells of $0.1 = \tilde{r}_s < \tilde{r} < 0.5$. In the last sections we concluded that the CPD in run C1 and C2 is mainly a complicated pass-through, with a very low permanent storage; it now becomes evident that some mass is accumulated in this region.

The only simulation that does not seem to accrete mass into the CPD is C1. This is interesting, as the flow pattern seen in this run is very similar to C2. The significant mass noise in this region makes it difficult to say whether C1 is not accreting or only slowly so.

The radial distributions of CPD surface densities, parameterised as $\Sigma(r)=\Sigma_0 r^{\,a}$, seem to be impacted only weakly by opacity effects, but dominated by the planetary potential depth. We use $\Sigma_0 = \Sigma(0.2\, r_{\rm H}) \approx 0.7\,\rm g\, cm^{-2}$, a value which is approximately the same for all runs at this radius. We find relatively steep power law-functions for the low-opacity runs C1 and C2, of $a = -2.5\pm0.2$ in the range $0.4<\tilde{r}<1.0$, that reflect the influence of the free-fall region on the mass profile, and a significantly flatter $a = -1.0\pm0.2$ in the range $0.1<\tilde{r}<0.3$, with some more fluctuations for C2 due to opacity changes; \cite{fujii2017} found similar deviations from the power-law behaviour, when using non-constant opacities. A slightly shallower distribution is found for run H1 with $a = -2.0\pm0.2$ over the whole envelope. The latter fits are, however, only extended over the relatively modest CPD region in those runs. 
A more realistic power-law is probably given by the CPD for run H2, for which we find a significantly flatter power-law with $a = -1.0\pm0.2$ for $0.1<\tilde{r}<1.0$ and in the pressure-influenced region $a = -2.0\pm0.2$ for $0.025<\tilde{r}<0.1$. 

The gas residence times in all CPDs are fairly low, hence the mass profiles are a direct consequence of the mode of radial mass transport that each individual CPD finds on this short timescale. A future set of simulations will have to be dedicated to study deep potentials like in H2, for a wider range of parameters. However, it is encouraging that the value for $a$ that we find in the CPD region of the nominal simulation run is extended over a much larger range in radii in run H2. Hence this is a hint that the true value for $a$ in our simulation settings is $a\approx-1$.

The runs with a surface density power law of $a\approx-2$ have most of their mass stored in the inner CPD, which is unresolved in those simulations. Run H2, which has a more realistic potential depth but is very hot, with $a\approx-1$, has its CPD mass $m\sim \int_{r_1}^{r_2} dr\,r\,\Sigma(r) \propto \Sigma_0 (r_2-r_1)$. Therefore the maximum CPD mass for this mass profile is $\Sigma_0 \,r_2 = \Sigma_0\, r_{\rm H}$. The value of $\Sigma_0$ is set by the mass flowing through the CPD, but it is conceivable that once the CPD cools down, the lifetime of the gas increases significantly in the CPD which would be stabilising the mass profile and allow for higher CPD accretion rates. 

Our measured CPD masses are very low compared to what is commonly assumed in models for the formation of moon systems  \citep{canup2006, cilibrasi2018, ronnet2020}. 
The ratio between our CPD masses and the central planetary mass is of the order $10^{-5}$.
\cite{isella2019} reported a CPD mass for PDS 70c of $\approx$100 times larger than our values. It is possible that this is a consequence of the surface density profile and the inner CPD truncation radius. However with the measured values for the CPD accretion rates (see Fig. \ref{fig:accretion_and_cpdmasses}) the needed dust mass can be delivered in $\approx $10\, kyrs, while for the Galilean satellites between 80 kyr and 0.32 Myr would be required.
    
\subsection{Gas accretion rates into the planetary envelopes}
\label{sec:gas_accretion_planets}

After discussing the CPD accretion rates, we now turn to analyse the envelope accretion rates. We list the measured CPD masses for the Jupiter-mass planets and the average accretion rates over the simulation time into the Hill-sphere for all our planets in Fig. \ref{fig:accretion_and_cpdmasses}. Our results show that for well-resolved simulations there is significant gas accretion, for all the considered opacities and for all planetary masses. This importantly implies that there is no problem in 3D gas accretion simulations once they are well-resolved, unlike previously stated in \cite{judith2014}. 

The accretion rate that we find for the 240$\rm m_{\oplus}$ planets for both opacities is slightly higher than for the 1$\rm m_{\rm J}$ planet, as can be seen in Fig. \ref{fig:accretion_and_cpdmasses}. From this we conclude that the Jupiter-mass planets are in a state of disc-limited accretion for the low opacity case, i.e. that the accretion rate is no longer limited by the Kelvin-Helmholtz contraction but rather by what the protoplanetary disc can provide through the Hill sphere. 
We refer to \cite{lambrechts2019} for further analysis of disc-limited versus cooling-limited accretion. This is classically understood to be a process in which the Kelvin-Helmholtz contraction of the envelope material is faster than any replenishment from the disc can happen. This results in the planet acting to accrete all the gas supplied through the Hill-sphere. The data points presented here differ quantitatively from the classic results in \cite{ayliffe2009a} mainly in that our accretion curve is flatter towards the low masses and we do not see a strong flattening of accretion rates for the high masses. Their results were obtained by using an SPH code, and it is possible that the differences to our $\dot{m}(m)$-curves result from differences in implementing radiation transport in eulerian vs. lagrangian codes.

In the context of the work  of \cite{cimerman2017}, it is perhaps surprising that we do not see any effects of entropy recycling at the low-mass end in our results, which would stop the gas accretion process.
Their simulations, while well-resolved, operate at lower planet masses (their maximum planet mass is 5$\rm m_{\oplus}$) and at much higher optical depths as well as disc masses (their $\rho_{\rm mid}\approx 6\times 10^{-6} \,\rm g \,cm^{-3}$, while we have $\rho_{\rm mid}\approx 4\times 10^{-11} \,\rm g \,cm^{-3}$ for the unperturbed discs of the 20$\rm m_{\oplus}$ planets). It is due to this main difference to their work that we do not see any important recycling effects for our low-mass planets.

\section{Summary}

We performed a set of 3D radiation hydrodynamics simulations in order to learn about the occurrence of circumplanetary discs and mass delivery onto them.

\begin{enumerate}
    \item Initially, we asked at which planet masses CPDs would occur and whether opacity is an important factor. We find that there is a significant shift in the rotation profiles of envelopes between 120$\rm m_{ \oplus}$ and 240$\rm m_{ \oplus}$, indicating the takeover of rotational support from pressure support. This indicates that CPDs form only late in a gas giants envelope accretion process. The residence time of gas in the CPDs is of the order of a few 10 CPD orbits; hence the rotation of the CPD gas is a direct result of newly accreted gas and not inherited from the initial conditions. We find that low opacity significantly helps flatten envelopes and reduce pressure support.  Surface density profiles in regions with significant rotational support stabilize to $\Sigma \sim r^{-1}$.
    \item Jupiter-mass planets accrete at disc-limited accretion rates of $10^{-2} \;\rm  m_{\oplus}\; yr^{-1}$. Mass enters the planetary Hill-sphere predominantly through the midplane and after being processed in the CPD ends up in the planet. This remains true even after extending the potential depth of our planets by a factor of four. CPDs seem to accrete with values of around $1-4\times 10^{-5} \; \rm m_{\oplus}\; yr^{-1}$, but those values are difficult to determine and there is a possibility that this is a numerical effect.
    \item The flows inside the CPD are influenced by the occurrence of a separate set of CPD spiral arms, as opposed to the classical CSD spiral arms, previously identified in the literature as tidal arms \citep{zhu2016}. Those CPD spiral arms cause gas to slowly rise above the midplane while it orbits, and finally to spiral into the planet at high altitude. The CPD spiral arms hence play a role in shaping the CPD structure and channeling the accreted gas into the planet.
    \item In our investigation of a low-opacity Jupiter (run C1), we have found a complete planetary detachment from the parent CSD, expressed as a region of pressure gradient inversion. We find a new circulation regime inside the Hill-sphere. In this case, the spiral arms feature low enough static pressure and, as a consequence, are overcome by the ram pressure of the accretion flow. This ram pressure evacuates a region between the spiral arm and the CSD.  
    This newly evacuated region facilitates free-fall onto the CPD, which causes a weak accretion shock in the midplane, additional to a pre-existing vertical accretion shock. The spiral arms are tilted through the overpressure, which can be used as a phenomenological identification mark for this effect. We note that this effect depends on the efficient radiative cooling of the spiral arm and functions at low envelope temperatures.
    \item Using low \cite{belllin1994} opacities (run C2) instead of constant ones reproduces a very similar simulation outcome to the one seen in C1.
    \item A high opacity (run H2) reproduces other work from the literature, with this envelope being an essentially featureless blob, with a large-scale circulation inside the Hill-sphere. Nonetheless the accretion rate for this run is on the order of $10^{-2} \; \rm m_{ \oplus}\; yr^{-1}$ into the smoothing length region.
    \item A simulation run with 4 times deeper potential, but otherwise identical parameters to C1, was investigated (run H1). We find very hot temperatures ($T_{\rm c}\approx$ 13000K), but significant rotation in the outer CPD ($v_{\theta}/v_{\rm K} \approx 0.8$) and an identical extent of the CPD to inner radii as in C1. This indicates the robustness of our findings about the CPD structure. Furthermore a lowered opacity, which can be obtained under a number of realistic circumstellar conditions \citep{ossenkopf1994} from various dust populations, seems to be paramount for forming CPDs.
\end{enumerate}

A major limitation of our model is nevertheless the limited numerical resolution of the inner regions of the Hill sphere. We have observed that increasing the resolution decreases the luminosity of the planet and affects the temperature and structure of the circumplanetary significantly. Therefore, future investigations of circumplanetary discs should be performed at higher resolutions in order to identify resolution convergence criteria regarding the structure of circumplanetary discs.

The constant turbulent viscosity of $\nu \approx 10^{15}\,\rm cm^2\, s^{-1}$ corresponding to $\alpha\approx 10^{-2}$ which we use, is high in the light of recent observational evidence \citep{pinte2016, dullemond2018}. Hence a re-examination of our results for lowered viscosities is in place. However, due to the significant runtimes of the gap formation at low viscosities in 3D RHD, we have to refer to future publications for this endeavour.


\begin{acknowledgements} We thank the anonymous referee for improving the quality of our manuscript and discussions of shock physics.
 MS was supported by a project grant from the Swedish Research Council (grant number 2014-5775). AJ wants to thank the support by the Knut and Alice Wallenberg Foundation (grant number 2017.0287), the European Research Council (ERC Consolidator Grant 724687-PLANETESYS) and VR grant 2018-04867. ML thanks the Knut and Alice Wallenberg Foundation under grant 2017.0287.
 BB thanks the European Research Council (ERC Starting Grant 757448-PAMDORA) for their financial support. All the simulations presented in this work were performed on resources provided by the Swedish National Infrastructure for Computing (SNIC) at Lunarc in Lund University, Sweden, and the entire team is grateful for being supported with their expertise.
\end{acknowledgements}

\begin{appendix}

\section{\textbf{On the origin of the spiral arm tilt}}

The vertically tilted spiral arms which we have documented, are the center point of the complex physics which we have analysed in this work. As their structure seems to be related to the cooling properties of the envelope and the disc, it is natural to pursue an explanation of the spiral arm tilt in the vertical temperature structure.

\begin{figure}
\centering
\includegraphics[width=0.5\textwidth]{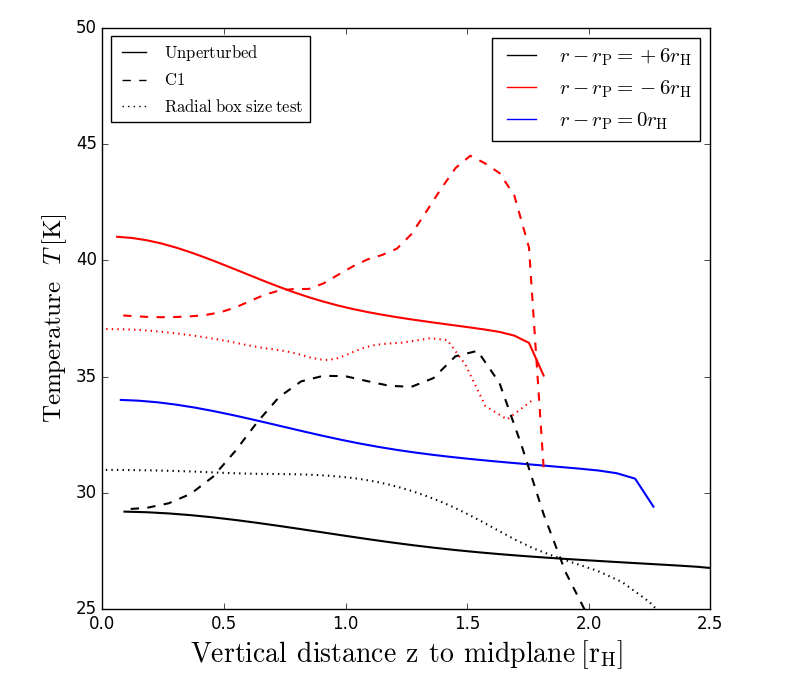}
\caption{Vertical, unperturbed temperature profile after 400 orbits before the planet is injected (solid red, blue and black lines), plotted against the distance to the midplane in units of the future planet. Radial positions of the  profiles are given in the plot, the azimuthal position is always $\theta=0$. The abrupt temperature decrease at the upper simulation boundary is due to the 3K boundary condition, but does not influence the simulation result because the disc there is optically thin. Dashed and dotted lines are after 5 orbits of the simulation C1, and a control simulation with larger radial box size. Blue for the $r=0$ in the highres runs is not plotted, as it appears in the paper in Figs. \ref{fig:c1_vertical_profiles} and \ref{fig:overview_opacities_profiles}. }
\label{fig:temperatureprofile}
\end{figure}

Therefore, in Fig. \ref{fig:temperatureprofile} we document the unperturbed temperature profiles in the disc at the position of the future planet and at the positions of the future gap edges. This simulation is run in 2D for 400 orbits in order to quickly find a radial-vertical equilibrium,. 
The temperature profiles are gaussian, as expected for a quasi-isothermal disc.

To put those into context, we also plot the temperatures for our nominal high-resolution simulation run C1 (dashed lines in Fig. \ref{fig:temperatureprofile}). Seeing the vertical temperature inversion, we assumed initially to have found the reason for the static pressure argument from Fig. \ref{fig:shock_mach_pressure}. However, a control simulation with larger radial box size (dotted lines in Fig. \ref{fig:temperatureprofile}) evolved to a different temperature structure with planet, due to a change in the flow in the upper disc layers (compressional heat plays an important role at those low temperatures). This temperature structure shows no temperature inversion, however the spiral arm tilt, with associated free-fall region was still evident in the control simulation.

This led us to reject the hypothesis that the spiral arm tilt is caused directly by the unperturbed disc vertical temperature gradients, and must lie with the self-consistently computed temperature solution in the envelope. We further note for possible future studies, that the spiral arm tilt is already evident in the gap-forming run. This should make it possible to find good candidate envelopes quickly if one wants to study the spiral arm tilt, without having the need to scan parameter space in expensive high resolution simulations.

\end{appendix}

\bibliographystyle{aa}
\bibliography{cpds} 

\end{document}